\documentclass[3p,final,times,sort&compress]{elsarticle}

\usepackage[utf8]{inputenc}
\usepackage[T1]{fontenc}
\usepackage{mathtools}
\usepackage{amsmath}
\usepackage{amsthm}
\usepackage{amssymb}
\usepackage{graphicx}
\usepackage{hyperref}
\usepackage{algorithm}
\usepackage{algpseudocode}
\usepackage{bm}
\usepackage[acronym,shortcuts]{glossaries}
\usepackage{graphicx}
\usepackage{subcaption}
\usepackage{xcolor}

\usepackage{etoolbox}
\AtBeginEnvironment{algorithmic}{\let\textbf\relax}

\newcommand{\vc}[1]{\mathbf{#1}}
\newcommand{\bs}[1]{\boldsymbol{#1}}
\newcommand{\dd}[0]{\,\mathrm{d}}
\newcommand{\pfrac}[2]{\frac{\partial #1}{\partial #2}}

\newcommand{\gain}[0]{K_{\tau\sigma}}
\newcommand{\gaino}[0]{K_{\tau\sigma,o}}
\newcommand{\gainh}[0]{K_{\tau\sigma,h}}
\newcommand{\stimw}[0]{\Delta\tau_w}
\newcommand{\stimi}[0]{\Delta\sigma_I}
\newcommand{\stimwh}[0]{\Delta\tau_{wh}}
\newcommand{\stimih}[0]{\Delta\sigma_{Ih}}
\newcommand{\comment}[1]{}

\newcommand{\dk}{\vc{d}_{t,k}}
\newcommand{\dtk}{\tilde{\vc{d}}_{t,k}}

\algblockdefx{Switch}{EndSwitch}[1]{\textbf{switch} (#1)}{\textbf{end switch}}
\algblockdefx{Case}{EndCase}[1]{\textbf{case} #1\textbf{:}}{\textbf{end case}}
\algblockdefx{Default}{EndDefault}[0]{\textbf{default:}}{\textbf{end case}}
\algdef{SE}[DOWHILE]{Do}{doWhile}{\algorithmicdo}[1]{\algorithmicwhile\ #1}%

\newacronym{3d}{3D}{three-di\-mensional}
\newacronym{cfd}{CFD}{computational fluid dynamics}
\newacronym{gr}{G\&R}{growth and remodeling}
\newacronym{fsi}{FSI}{fluid-structure interaction}
\newacronym{fsg}{FSG}{fluid-solid-growth interaction}
\newacronym{fsge}{FSGe}{equilibrated fluid-solid-growth}
\newacronym{cmm}{CMM}{constrained mixture model}
\newacronym{cmme}{CMMe}{equilibrated constrained mixture model}
\newacronym{rbcmm}{rb-CMM}{rate-based constrained mixture model}
\newacronym{wss}{WSS}{wall shear stress}
\newacronym{ims}{IMS}{intramural stress}

\newcommand{\reva}[1]{\textcolor{black}{#1}}
\newcommand{\revb}[1]{\textcolor{black}{#1}}
\newcommand{\revc}[1]{\textcolor{black}{#1}}

\journal{Computer Methods in Applied Mechanics and Engineering}

\begin{document}
\begin{frontmatter}

\title{FSGe: A fast and strongly-coupled 3D fluid-solid-growth interaction method}

\author[inst1]{Martin~R. Pfaller}
\author[inst2]{Marcos Latorre}
\author[inst3,inst4]{Erica~L. Schwarz}
\author[inst1]{Fannie~M. Gerosa}
\author[inst1]{Jason~M. Szafron}
\author[inst4]{Jay~D. Humphrey}
\author[inst1]{Alison~L. Marsden}

\affiliation[inst1]{
organization={Department of Pediatrics -- Cardiology, Stanford Univeristy},
city={Stanford},
state={CA 94306},
country={USA}}

\affiliation[inst2]{
organization={Center for Research and Innovation in Bioengineering, Universitat Politècnica de València},
city={València},
country={Spain}}

\affiliation[inst3]{
organization={Department of Bioengineering, Stanford Univeristy},
city={Stanford},
state={CA 94306},
country={USA}}

\affiliation[inst4]{
organization={Department of Biomedical Engineering, Yale Univeristy},
city={New Haven},
state={CT 06520},
country={USA}}

\begin{abstract}
\Gls*{fsge} is a fast, open source, \gls*{3d} computational platform for simulating interactions between instantaneous hemodynamics and long-term vessel wall adaptation through \revc{mechanobiologically equilibrated} \gls*{gr}. Such models \revc{can capture evolving geometry, composition, and material properties} in health and disease and following clinical interventions. In traditional \gls*{gr} models, this feedback is modeled through highly simplified fluid \revc{solutions}, neglecting local variations in blood pressure and \gls*{wss}. FSGe overcomes these inherent limitations by strongly coupling the \gls*{3d} Navier-Stokes equations for blood flow with a \gls*{3d} \gls*{cmme} for vascular tissue \gls*{gr}. \gls*{cmme} allows one to predict long-term evolved mechanobiological equilibria from an original homeostatic state at a computational cost equivalent to that of a standard hyperelastic material model. In illustrative computational examples, we focus on the development of a stable aortic aneurysm in a mouse model to highlight key differences in growth patterns between \gls*{fsge} and solid-only \gls*{gr} models. We show that \gls*{fsge} is especially important in blood vessels with asymmetric stimuli. Simulation results reveal greater local variation in fluid-derived \gls*{wss} than in \gls*{ims}. Thus, differences between \gls*{fsge} and \gls*{gr} models became more pronounced with the growing influence of \gls*{wss} relative to pressure. Future applications in highly localized disease processes, such as for lesion formation in atherosclerosis, can now include spatial and temporal variations of \gls*{wss}.
\end{abstract}




\end{frontmatter}

\section{Introduction\label{sec_intro}}
\begin{figure}[H]
\centering
\includegraphics[width=.7\linewidth]{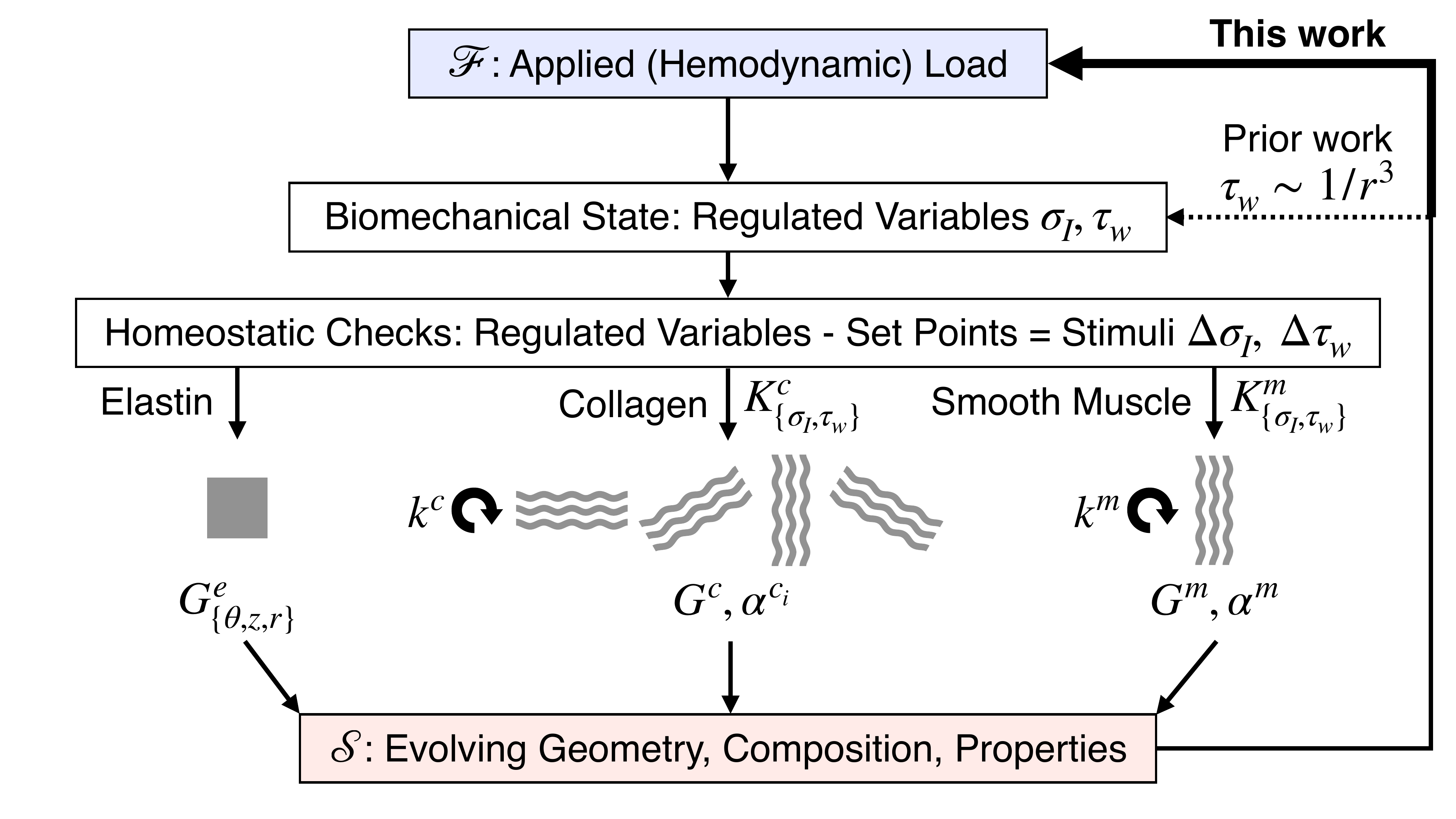}
\caption{Negative feedback characteristics of mechanical homeostasis, including hemodynamics $\mathcal{F}$ and solid mechanics $\mathcal{S}$ . Stimuli consist of deviations of intramural stress, $\stimi$, and wall shear stress, $\stimw$, from homeostatic set-point values. Tissue changes are modulated by gain factors $K^\alpha_{\sigma_I}$ and $K^\alpha_{\tau_w}$, which capture cell sensitivity to the particular stimulus. \Gls*{gr} are influenced further by degradation rate $k^\alpha$, homeostatic prestretch $G^\alpha$, and orientation angle $\alpha^\alpha$. In this work, tissue constituents $\alpha$ are elastin $e$, four collagen fiber families $c$, and smooth muscle $m$, which are visualized in Figure~\ref{fig_constituents}. Figure adapted from \cite{humphrey21}.\label{fig_fsg}}
\end{figure}

The intricate interaction between hemodynamics and growth (change in mass) and remodeling (change in microstructure) of a blood vessel plays a vital role in development, homeostasis, and disease progression \cite{loerakker22}. The \gls*{cmm} simulates the continuous deposition and degradation of tissue constituents \cite{humphrey02,humphrey21}, often based on the concept of mechanical homeostasis, which is visualized in Figure~\ref{fig_fsg}. In this model of mechanobiological homeostasis, changes in vascular composition and properties stem from deviations in regulated variables from defined set points. These regulated variables primarily include \revb{scalar metric measures of} \gls*{ims}, $\sigma_I$, and \gls*{wss}, $\tau_w$, which have consistently been found to be maintained \revc{near} homeostatic values under physiologic conditions \textit{in vivo} \cite{dajnowiec07}.

Hemodynamics are a crucial determinant of the biomechanical state of a blood vessel, characterized in part by its regulated variables. Solid mechanics within the vessel wall define blood vessel geometry and microstructure. Changes in the solid mechanics thus trigger changes in hemodynamics rendering \revc{this} feedback a \gls*{fsg} problem. Most prior work considered either constant or reduced-order hemodynamics. The novelty of our work is a \textit{strongly} coupled feedback between hemodynamics and solid mechanics in a fully \gls*{3d} framework based on the \revc{mechanobiologically} \gls*{cmme} \cite{latorre20}. In the following, we review previous approaches in more detail and highlight the contributions of this work.

Hemodynamics \textit{in vivo} typically include complex flow phenomena such as separation and recirculation induced by changes in cross-sectional area or bifurcations \cite{secomb16}. Consequently, pressure and \gls*{wss} typically show a high spatial and temporal variation. Local hemodynamics can be quantified with \gls*{cfd} or \gls*{fsi} solutions \cite{schwarz23}. However, the \gls*{cmm} commonly includes simplified assumptions for axial pressure gradient $\Delta p$ and \gls*{wss} $\tau_w$ based on Hagen-Poiseuille law \cite{latorre20}
\begin{equation}
\Delta p = \frac{8\mu l Q}{\pi \revc{a}^4}, \quad \tau_w = \frac{4 \mu Q}{\pi \revc{a}^3},
\label{eq_poiseuille}
\end{equation}
with dynamic blood viscosity $\mu$, vessel length $l$ and inner radius $a$, and bulk blood flow rate $Q$. These equations assume fully developed steady-state flow through a straight cylindrical tube with axisymmetric flow and zero radial and circumferential components, which is generally not upheld in \textit{in vivo} settings.

The local interaction between hemodynamics and vessel \gls*{gr} mechanisms is crucial in numerous diseases and their treatments, as, for example, in aneurysms \cite{watton09,aparicio14,selimovic14,teixeira20}, tissue-engineered vascular grafts \cite{schwarz21,blum22}, vein graft adaptation and failure following coronary artery bypass graft surgery \cite{ramachandra16,khan20}, and in congenital heart patients with pulmonary disease \cite{lan22,szafron23}. For an overview of \gls*{gr} models, i.e., kinematic \reva{\cite{rodriguez94}}, \gls*{cmm} \reva{\cite{humphrey02}}, and \reva{\gls*{rbcmm} \cite{watton04}}, see \cite{ambrosi19}. A non-comprehensive overview of similar fluid-solid-growth interaction models is given in Table~\ref{tab_fsg}.

\revc{Baek et al. \cite{humphrey07,baek07b} devised a simple control-volume network model for the flow that was coupled to a membrane \gls*{cmm}.} Figueroa et al. \cite{figueroa09} developed a method that couples \gls*{3d} hemodynamics to a membrane model of the vessel wall. This theory is based on the \textit{theory of small on large} \cite{baek07}, superimposing "small" deformations during the cardiac cycle upon "large" deformations due to \gls*{gr}. The membrane wall formulation is related to the coupled momentum method for \gls*{fsi} in blood vessels \cite{figueroa06}. The long-term \gls*{gr} deformation is coupled to pulsatile hemodynamics, illustrated in an example of aneurysmal development. This approach was used in a patient-specific aneurysm geometry in Sheidaei et al. \cite{sheidaei11}.

\reva{Watton et al. (2009) \cite{watton09} integrated the membrane model from Watton et al. (2004) \cite{watton04} into an \gls*{fsg} framework, simulating saccular aneurysms on idealized 3D arterial geometries. Aparicio et al. \cite{aparicio14} considered cyclic deformations as a \gls*{gr} stimulus for collagen with a spatially and temporally heterogeneous endothelium. Selimovic et al. \cite{selimovic14} showed that spatially heterogeneous \gls*{wss} can account for asymmetry in intracranial aneurysms. Teixeria et al. \cite{teixeira20} considered the wall shear stress aspect ratio \cite{vamsikrishna20} as a \gls*{gr} stimulus, a metric for oscillatory flow.}

The work of Latorre et al. \cite{latorre22} couples 3D \gls*{gr} with a simplified control-volume formulation of fluid dynamics. Pressure and \gls*{wss} are derived from an axial node-based control-volume approach based on the Hagen-Poiseuille law \eqref{eq_poiseuille} with additional pressure loss terms. For a more detailed discussion of reduced-order fluid dynamics, see Pfaller et al. \cite{pfaller22}. Most recently, Schwarz et al. \cite{schwarz23b} coupled \gls*{3d} \gls*{cmm} and \gls*{3d} hemodynamics. While this constitutes the most detailed approach so far, it comes at high computational cost due to the need to integrate heredity integrals point-wise in full \gls*{cmm}.

Most FSG approaches reviewed above, except for \revc{Baek et al. \cite{baek07b}}, Latorre et al. \cite{latorre22}, and Schwarz et al. \cite{schwarz23b}, employ weak coupling between fluid and solid \gls*{gr}. That is, the fluid and solid domains are solved independently, with their solutions being exchanged only once in each load step. This approach assumes that the interaction between the fluid and solid is relatively weak, allowing for some degree of decoupling between the domains. Loose coupling is typically used for problems where the fluid and solid have a minor influence on each other or when computational efficiency is prioritized over solution accuracy. On the other hand, strong coupling schemes rely on an iterative procedure that ensures the convergence of fluid and solid solutions.

\begin{table}[htbp]
\centering
\begin{tabular}{|llllll|}
\hline
\textbf{Reference} & \textbf{Fluid} & \textbf{Solid} & \textbf{Stimuli} & \textbf{FS-Coupling} & \textbf{Geometry} \\
\hline
\revc{Baek \cite{humphrey07,baek07b}} & \revc{Control-volume} & \revc{membrane CMM} & \revc{pulsatile mean} & \revc{strong} & \revc{idealized} \\
Figueroa \cite{figueroa09} & \gls*{3d} Navier-Stokes & membrane CMM & pulsatile mean & weak & idealized \\
Sheidaei \cite{sheidaei11} & \gls*{3d} Navier-Stokes & membrane CMM & pulsatile mean & weak & \textit{in vivo} \\
Watton \cite{watton09} & \gls*{3d} Navier-Stokes & \reva{membrane rb-CMM} & pulsatile mean & weak & \reva{idealized} \\
\reva{Aparicio \cite{aparicio14}} & \reva{\gls*{3d} Navier-Stokes} & \reva{membrane rb-CMM} & \reva{pulsatile mean, cyclic} & \reva{weak} & \reva{idealized}\\
\reva{Selimovic \cite{selimovic14}} & \reva{\gls*{3d} Navier-Stokes} & \reva{membrane rb-CMM} & \reva{steady-state} & \reva{weak} & \reva{\textit{in vivo}}\\
Grytsan \cite{grytsan15} & \gls*{3d} Navier-Stokes & 3D \reva{rb-CMM} & steady-state & weak & \textit{in vivo} \\
\reva{Teixeria \cite{teixeira20}} & \reva{\gls*{3d} Navier-Stokes} & \reva{3D rb-CMM} & \reva{pulsatile \& oscillatory} & \reva{weak} & \reva{\textit{in vivo}}\\
Latorre \cite{latorre22} & Control-volume & \gls*{3d} \gls*{cmme} & steady-state & strong & idealized \\
Schwarz \cite{schwarz23b} & \gls*{3d} Navier-Stokes & \gls*{3d} \gls*{cmm} & steady-state & strong & \textit{in vivo} \\
\textit{This work} & \gls*{3d} Navier-Stokes & \gls*{3d} \gls*{cmme} & steady-state & strong & idealized \\
\hline
\end{tabular}
\caption{\glsresetall Overview of some \gls*{fsg} interaction models.\label{tab_fsg}}
\end{table}

This work introduces an \gls*{fsge} method to strongly couple 3D \gls*{gr} based on the \gls*{cmme} with 3D hemodynamics. Using \gls*{cmme} greatly reduces the computational cost of evaluating the solid model and generally arrives at the desired model prediction in much fewer steps. In the following, we introduce the governing differential equations of the fluid and solid domains (Section~\ref{sec_equaitions}). We then review computational techniques employed to solve the coupled fluid-solid problem (Section~\ref{sec_fsge}). In illustrative computational examples of an asymmetric aneurysm, we compare key differences between a solid-only \gls*{gr} model and our \gls*{fsge} formulation (Section~\ref{sec_results}). We close with a discussion of the results, limitations, and future perspectives (Section~\ref{sec_discussion}).

\section{Governing equations of fluid and solid domains\label{sec_equaitions}}
%
As outlined in  Figure~\ref{fig_fsg}, the \gls*{fsge} model relies on the solution of fluid (\ref{sec_eqs_fluid}) and solid (\ref{sec_eqs_solid}) mechanics, whose governing equations we briefly review in this section. Additionally, for completeness we introduce both the full \gls*{cmm} (\ref{sec_cmm}) and \gls*{cmme} (\ref{sec_cmm_eq}) \gls*{gr} models. Both \gls*{3d} fluid and solid were solved numerically in our open-source multi-physics solver \texttt{svFSIplus} \cite{svFSIplus}, the C\texttt{++} version of \texttt{svFSI} \cite{zhu22b}; both solvers are released with the SimVascular project \cite{updegrove16}.

\subsection{Fluid dynamics\label{sec_eqs_fluid}}
We model blood flow in large vessels at high shear rates as a Newtonian fluid in the quasi-static fluid domain $\Omega^\text{f}$ within time scale $T^\text{f}$ with the incompressible Navier-Stokes equations, 
\begin{alignat}{3}
\nabla \cdot \boldsymbol{\sigma}^\text{f}(\bm{u}, p) &= \rho \left[ \frac{\partial\bm{u}}{\partial t} + (\bm{u} \cdot \nabla) \bm{u} \right] , \quad && \bm{x} \in \Omega^\text{f}, \quad t \in [0,T^\text{f}],\\
\nabla \cdot \bm{u} &= 0, \quad && \bm{x} \in \Omega^\text{f}, \quad t \in [0,T^\text{f}],
\label{eq_navier_stokes}
\end{alignat}
with density $\rho$, viscosity $\mu$, and fluid Cauchy stress tensor $\boldsymbol{\sigma}^\text{f} = \mu(\nabla\bm{u} + \nabla\bm{u}^\intercal)-p\bm{I}$, where $\bm{u}$ and $p$ are fluid velocity and pressure, respectively. We prescribe boundary conditions
\begin{alignat*}{4}
\bm{u}(\bm{x}, t) &= \vc{0}, \qquad && \bm{x} \in \Gamma_\text{int}, \quad && t \in [0,T^\text{f}],\\
\bm{u}(\bm{x}, t) &= \bm{u}_\text{in}(\bm{x}), \qquad && \bm{x} \in \Gamma_\text{f,in}, \quad && t \in [0,T^\text{f}],\\
\bm{n} \cdot \boldsymbol{\sigma}^\text{f}(\bm{x}, t) \cdot \bm{n} &= p_\text{out}, \qquad && \bm{x} \in \Gamma_\text{f,out}, \quad && t \in [0,T^\text{f}],
\end{alignat*}
with a no-slip boundary condition at the fluid-solid interface $\Gamma_\text{int}$, constant quadratic inflow profile $\bm{u}_\text{in}$ at the inlet $\Gamma_\text{f,in}$, and constant outlet pressure $p_\text{out}$ at the outlet $\Gamma_\text{f,out}$ with normal $\bm{n}$. The initial conditions are
\begin{alignat*}{3}
\bm{u}(\bm{x}, t=0) &= \bm{u}_0(\bm{x}), \qquad && \bm{x} \in \Omega^\text{f},\\
p(\bm{x}, t=0) &= p_0(\bm{x}), \qquad && \bm{x} \in \Omega^\text{f},
\end{alignat*}
with initial velocity field $\bm{u}_0$ and initial pressure field $p_0$. See Esmaily Moghadam et~al. \cite{moghadam13,esmailymoghadam13} for details on the P1-P1 variational multiscale finite element solution of the Navier-Stokes equations.

\subsection{Solid growth and remodeling\label{sec_eqs_solid}}
We follow the classic approach of finite strain theory to model the vessel wall displacement, $\bm{d} = \bm{x}-\bm{X}$, with reference position $\bm{X}$ and current position $\bm{x}$. We calculate the deformation gradient $\bm{F}$, the Jacobian $J$, and the right Cauchy-Green tensor $\bm{C}$ as
\begin{equation*}
\bm{F} = \pfrac{\bm{x}}{\bm{X}}, \quad J=\text{det}\, \bm{F}, \quad \bm{C} = \bm{F}^\text{T}\bm{F}.
\end{equation*}
The balance of linear momentum in the case of negligible \revc{body forces and }inertial loads \cite{humphrey02c} in the solid domain $\Omega^\text{s}$ yields the quasi-static problem
\begin{alignat}{3}
\nabla \cdot \bs{\sigma}^\text{s} = \bm{0}, \qquad && \bm{x} \in \Omega^\text{s}, \quad && t \in [0,T^\text{s}],
\label{eq_solid}
\end{alignat}
with solid stress tensor $\bs{\sigma}^\text{s}$ provided constitutively by the \gls*{gr} \revc{constitutive relation} and its solid time scale $T^\text{s}$. We prescribe boundary conditions
\begin{alignat*}{4}
d_a(\bm{x}, t) &= 0, \qquad && \bm{x} \in \Gamma_\text{cap}, \quad && t \in [0,T^\text{s}],\\
\bs{\sigma}^\text{s}(\bm{x}, t) \cdot \bm{n}(\bm{x}, t) &= k \, \bm{d}(\bm{x}, t), \qquad && \bm{x} \in \Gamma_\text{out}, \quad && t \in [0,T^\text{s}],
\end{alignat*}
with zero Dirichlet boundary conditions in axial direction $z$ at the caps of the blood vessel $\Gamma_\text{cap}$ and external tissue support with outward normal $\bm{n}$ and stiffness $k$ \cite{moireau11}. Note that our external tissue support does not include viscous damping since \gls*{gr} is a quasi-static process. The initial condition is
\begin{alignat*}{3}
\bm{d}(\bm{x}, t=0) &= \bm{0}, \qquad && \bm{x} \in \Omega^\text{s},
\end{alignat*}
We use a \gls*{gr} constitutive law to relate deformation $\bm{F}$ to stresses $\bs{\sigma}^\text{s}$. In the following, we briefly review the full \gls*{cmm} and the \gls*{cmme}. The \gls*{cmm} has been diversely applied, e.g., in altered flow and pressure conditions \cite{valentin08}, abdominal \cite{wilson12} and thoracic \cite{latorre20b} aortic aneurysms, vein grafts \cite{ramachandra20}, or tissue-engineered vascular grafts \cite{szafron18}. However, due to its computational complexity, there are only limited implementations in three dimensions \cite{valentin13,horvat19}. In this work, we use the fast, rate-independent \gls*{cmme} method that directly predicts the long-term evolved solution that is in mechanobiological equilibrium \cite{latorre18b}. This material model had been previously implemented in the open-source finite element solver \texttt{FEBio} \cite{maas12}, with computational costs comparable to hyperelastic materials \cite{latorre20}. Prior work relied on constant-pressure models \cite{latorre20,latorre20b,goswami22} and control-volume hemodynamics \cite{latorre22}. We briefly introduce the \gls*{cmm} in Section~\ref{sec_cmm} and outline the \gls*{cmme} in Section~\ref{sec_cmm_eq} but refer interested readers to the original publications.

\begin{figure}
\centering
\includegraphics[width=.5\linewidth]{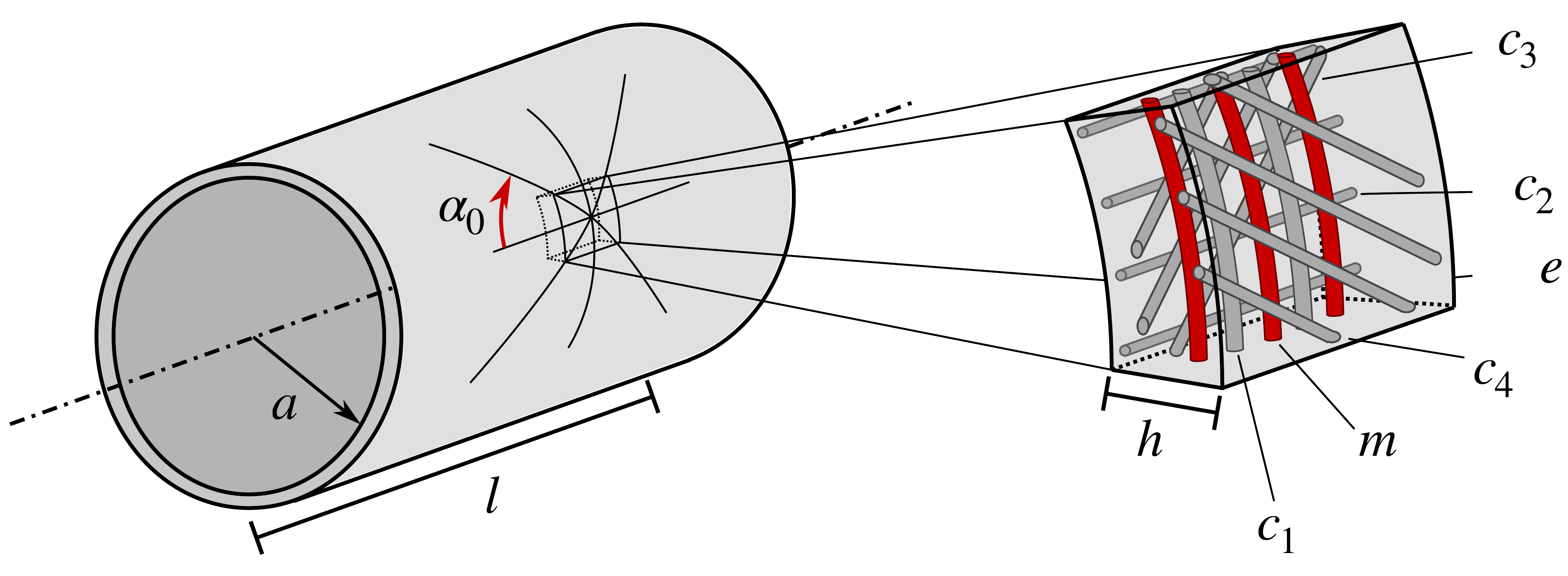}
\caption{Blood vessel with length $l$, \revc{inner radius $a$}, and thickness $h$. Its constituents are elastin $e$ (isotropic), smooth muscle $m$ (red, circumferential), and four collagen fiber families $c_i$ (circumferential, axial, diagonal angle $\alpha_0$). Image created by Sebastian L. Fuchs and licensed under the Creative Commons Attribution 4.0 International License.\label{fig_constituents}}
\end{figure}

\subsubsection{Constrained Mixture Model\label{sec_cmm}}
The \gls*{cmm} assumes that structurally significant constituents $\alpha$ with individual reference configurations, material properties, mass balance relations, and turnover rates are locally constrained to move together. The mass production per unit volume at  \gls{gr} time $\tau\in[0,s]$, where $s$ is the current \gls*{gr} time and $\tau$ is an intermediate time at which constituents are deposited, is expressed as
\begin{align*}
m^\alpha(\tau) = m^\alpha_o \, \Upsilon^\alpha(\tau),
\end{align*}
using an original rate $m^\alpha_o>0$ that is modulated by a stimulus function, as, for example,
\begin{align}
\Upsilon^\alpha(\tau) = 1 + K^\alpha_\sigma \, \stimi(\tau) - K^\alpha_\tau \, \stimw (\tau),
\label{eq_stimulus}
\end{align}
with gain parameters $K^\alpha_i>0$ controlling the sensitivity to stress deviations $\stimi$ and $\stimw$, respectively. We define these stimuli as
\begin{align}
\stimi(\tau) = \frac{\sigma_{I}(\tau)}{\sigma_{Io}} - 1, \quad \stimw(\tau) = \frac{\tau_{w}(\tau)}{\tau_{wo}} - 1,
\label{eq_stimuli_transient}
\end{align}
governed by changes in \revb{the scalar metrics} \gls*{ims} $\sigma_I$ (induced by pressure and axial force) and \gls*{wss} $\tau_W$ (induced by blood flow) from homeostatic set points $\sigma_{Io}$ and $\tau_{wo}$, respectively. The constituent removal between deposition times $\tau$ and current time $s$ is modeled by an exponential decay
\begin{equation*}
q^{\alpha}(s,\tau) = \exp \left( - \int_\tau^s k^\alpha(t) \dd t\right), \quad k^\alpha(t) = k^\alpha_o \, \left[1 + K^\alpha \, \Delta\sigma_I^2(t)\right],
\end{equation*}
where $k^\alpha$ is a rate parameter with homeostatic value $k^\alpha_o$, and $K^\alpha > 0$ is a gain parameter for \gls*{ims} deviation. We can express the mass density per current volume of each constituent $\rho^\alpha$ by
\begin{equation}
\rho^\alpha(s)=\int_{-\infty}^{s} m^\alpha(\tau) \, q^{\alpha}(s,\tau) \dd\tau.
\label{eq_cmm_int}
\end{equation}
The constituent-specific elastic strain energy density functions $\hat{W}^\alpha$ for elastin $e$, collagen $c$, and smooth muscle $m$ are characterized by the neo-Hookean and Fung-type exponential relations as
\begin{align}
\hat{W}^{e} = c^e (I^e_{1} - 3), \quad 
\hat{W}^{m,c} = \frac{c_1^{m,c}}{4c_2^{m,c}} \left[ \exp \left(c_2^{m,c}  (I_4^{m,c} - 1)^2 \right) - 1 \right],
\label{eq_hyperelastic}
\end{align}
where $c^e,c_1^{m,c},c_2^{m,c}$ are material parameters and $I^e_{1},I_4^{m,c}$ are constituent-specific measures of deformation that are invariant to the coordinate system used. The constituents are visualized in Figure~\ref{fig_constituents}. Collagen fibers are further classified into circumferentially, axially, and diagonally oriented (at angle $\alpha_0$) populations with fractions $\beta_\theta$, $\beta_z$, and $\beta_d$, respectively. The constituent relations \eqref{eq_hyperelastic} are evaluated as $\hat{W}^\alpha(\bm{F}^\alpha_{n(\tau)}(s))$ with the deformation gradient
\begin{align}
\bm{F}^\alpha_{n(\tau)}(s) = \bm{F}(s) \, \bm{F}^{-1}(\tau) \, \bm{G}^\alpha(\tau),
\label{eq_homeostasis}
\end{align}
where $\bm{G}^\alpha$ is the constituent-specific preferred homeostatic pre-stretch, which can be reduced to the scalars $G^{m,c}$ for smooth muscle cells and collagen fibers. The elastin prestretch tensor $\bm{G}^e$ is characterized by the components $G^e_{\theta,r,z}$. We obtain the total strain energy density
\begin{align*}
W(s) = \sum_\alpha W^\alpha(s), \quad W^\alpha(s) = \int_{-\infty}^{s} \frac{m^\alpha(\tau) \, q^{\alpha}(s,\tau)}{\rho(s)} \, \hat{W}^{\alpha}_{n(\tau)} \left(\bm{F}^\alpha_{n(\tau)}(s) \right) \dd\tau.
\end{align*}
The constituent-specific mass fraction $\phi^\alpha$ at time $s$ is defined as
\begin{align*}
\phi^\alpha(s) = \frac{\rho^\alpha(s)}{\rho(s)}.
\end{align*}

\subsubsection{Equilibrated Constrained Mixture Model\label{sec_cmm_eq}}
The \gls*{cmme} directly predicts the long-term evolved state in mechanical and biological equilibrium, i.e., at $s\to\infty$, where production again balances removal. Note that \gls*{cmme} can still be used to predict transient \gls*{gr} phenomena as long as the process is considered to be quasi-equilibrated, similar to classical problems in mechanics of quasi-equilibrated problems. Toward this end, it is assumed that the imposed perturbations to the homeostatic state are slow relative to the adaptive \gls*{gr} process. For this purpose, the \gls*{gr} process is broken up into several quasi-static \textit{load steps} \revc{$t$} where perturbations are applied in increments\revc{, as opposed to a \textit{time step} in a dynamic problem. All load steps represent a long-term equilibrium at $s\to\infty$, typically for different insult severities.} This rate-independent approach does not require computing hereditary integrals \eqref{eq_cmm_int}, thus drastically lowering the computational demand. In \gls*{cmme}, the stimulus function in Equation~\eqref{eq_stimulus} reduces to
\begin{align}
\Upsilon^\alpha_h (\Delta\sigma_{Ih}, \Delta\tau_{wh}) = 1,
\end{align}
depending on the equilibrated stimulus functions that can be simplified from \eqref{eq_stimuli_transient} as
\begin{align}
\stimih = \frac{\sigma_{Ih}}{\sigma_{Io}} - 1, \quad \stimwh = \frac{\tau_{wh}}{\tau_{wo}} - 1,
\label{eq_stimuli_equilibrated}
\end{align}
comparing the evolved equilibrium $h$ to the original homeostasis $o$, hence eliminating the time-dependency on $\tau$. All values with index $o$ are determined at a hyperelastic pre-loading stage ($t=0$), followed by the \gls*{gr} stage $h$ to determine the long-term evolved variables in mechanobiological equilibrium ($t>0$). The equilibrated stimulus function yields
\begin{align}
\Delta\sigma_{Ih} = \gain \, \Delta\tau_{wh}, \quad \gain = \frac{K^\alpha_\tau}{K^\alpha_\sigma}\geq0~\forall\alpha, \quad \eta = \frac{K^m}{K^c},
\label{eq_stimulus_eq}
\end{align}
with a single shear-to-intramural gain ratio $\gain$ for all constituents $\alpha$ and smooth muscle-to-collagen turnover ratio $\eta$. Note that for $\gain=0$ and $\gain\to\infty$ the blood vessel perfectly maintains \gls*{ims} $\sigma_{Ih}$ and \gls*{wss} $\tau_{wh}$, respectively. We obtain the solid Cauchy stress
\begin{align}
\bs\sigma^s = \bs\sigma^x - p_h \bm{I}, \quad \bs\sigma^x = \sum_\alpha^{e,c,m} \phi_h^\alpha \hat{\bs{\sigma}}_h^\alpha,
\label{eq_cauchy_stress}
\end{align}
where $\hat{\bs{\sigma}}^\alpha_h$ are the equilibrated constituent-specific Cauchy stresses, $\bs\sigma^x$ is the \revc{extra part of the} mixture stress, and $p_h$ is the equilibrated Lagrange multiplier that enforces mechanobiological equilibrium. We define the \gls*{ims} invariant throughout this work as
\begin{align}
\sigma_I = \frac{\text{tr}\,\bs\sigma^s}{3},
\label{eq_sigma}
\end{align}
calculated from the solid stress tensor $\bs\sigma^s$. We can now express the Lagrange multiplier, using \eqref{eq_stimulus_eq}, \eqref{eq_cauchy_stress}, and \eqref{eq_sigma}, as
\begin{align}
p_h = \sigma^x_{Ih} - \sigma_{Io} \left[1+K_{\tau\sigma} \left(\frac{\tau_{wh}}{\tau_{wo}} - 1 \right)\right], \quad \sigma^x_{Ih} = \frac{\text{tr}\,\bs\sigma^x}{3}.
\end{align}
The invariant $\sigma^x_{Ih}$ can be explicitly calculated from the mixture stress $\bs\sigma^x$. The original homeostatic \gls*{ims} $\sigma_{Io}$ is stored locally from the pre-loading stage. Note that the tangent stiffness matrix of the \gls*{cmme} material model only has minor symmetries, no major symmetries. This can pose challenges for iterative linear solvers and their preconditioners. We thus use the distributed memory sparse direct linear solver \texttt{MUMPS} \cite{amestoy01b} within the \texttt{PETSc} suite \cite{balay19}.

\section{Fluid-solid-growth coupling\label{sec_fsge}}
%
In classic \gls*{fsi}, the solid material behavior is determined by an \textit{elastic} response to boundary conditions and interface conditions at the fluid-solid interface $\Gamma_\text{int}$. In contrast, the \gls*{cmme} material behavior is characterized by a \textit{long-term} mechanobiological response to mechanical stimuli. In this section, we first highlight key similarities and differences between \gls*{fsi} and \gls*{fsge} problems and then derive a robust coupling scheme. \Gls*{wss} is a crucial stimulus in the models used in this work. Its local calculation is a key difference between purely solid \gls*{gr} models and \gls*{fsge}:
\begin{align}
\text{\gls*{gr}:}~\stimwh &\approx \left( \frac{a_h}{a_o} \right)^{-3} - 1 \approx \left[ \frac{r_o}{a_o} \, \lambda_\theta - \left(\frac{r_o}{a_o} - 1\right) \, \lambda_r \right]^{-3} - 1, \quad \lambda_i = \sqrt{\vc{C}_h : \vc{e}_i \otimes \vc{e}_i}.\\
\text{\gls*{fsge}:}~\stimwh &= \frac{\Vert \bs{\tau}_{wh} \Vert}{\Vert \bs{\tau}_{wo} \Vert} - 1.
\label{eq_stimuli}
\end{align}
In the \gls*{gr} model, local \gls*{wss} is approximated with the Poiseuille flow-derived \gls*{wss} \eqref{eq_poiseuille}. The evolved local inner radius $a_h$ of the blood vessel is approximated locally at each Gau\ss~point of the finite elements. The approximation uses the radial coordinate $r_o$ and inner radius $a_o$ in original homeostasis as well as the evolved local circumferential stretch $\lambda_\theta$ and radial stretch $\lambda_r$ with unit vectors $\vc{e}_i$ \cite{latorre20}. In the \gls*{fsge} model, we use the magnitude of the local \gls*{wss} vector $\bs{\tau}_w$ extracted at the fluid-solid interface from a steady-state fluid solution. In response to local changes in \gls*{wss}, the endothelium alters its production of vasoactive molecules that rapidly diffuse into the wall and cause vascular smooth muscle cell relaxation or contraction \cite{humphrey15}. Instead of solving the diffusion equation, this work assumes that the \gls*{wss} stimulus at the interface $\Gamma_\text{int}$ affects each radial location equally within the vessel wall. Thus, the \gls*{fsge} model is coupled via
\begin{alignat}{3}
\bs{\sigma}^\text{s} \vcentcolon= \bs{\sigma}^\text{s}(\bs{\tau}_w) = \bs{\sigma}^\text{s}(\bs{\sigma}^\text{f}_{\Gamma_\text{int}}), \qquad \bm{x} \in \Omega^\text{s},
\end{alignat}
to the solid stress tensor $\bs{\sigma}^\text{s}$. By modeling the effect of vasoactive molecules at every location in the vessel, the solid stress $\bs{\sigma}^\text{s}$ depends on \gls*{wss} $\bs{\tau}_w$, which, in turn, depends on the fluid stress tensor $\bs{\sigma}^\text{f}_{\Gamma_\text{int}}$ at the interface $\Gamma_\text{int}$. This interaction differs from classic \gls*{fsi} problems, where fluid and solid are only coupled at the interface. In \gls*{fsge}, the fluid-solid interface is coupled to the whole solid volume. Furthermore, in blood flow, \gls*{wss} $\tau_w$ is typically five orders of magnitude smaller than the \gls*{ims} $\sigma_I$. Thus, \gls*{wss} has a negligible \textit{mechanical} influence in vascular \gls*{fsi} models, but, depending on the gain ratio $\gain$, a large \textit{biological} influence in the \gls*{fsge} model. Also, solid \gls*{gr} takes place on a time scale $T^\text{s}\gg T^\text{f}$ of weeks to months, which is much larger than the fluid dynamics time scale $T^\text{f}$ of milliseconds to seconds. Thus, we can derive the interface velocity from the interface displacements $\bm{d}$ as
\begin{alignat}{3}
\bm{u}_{\Gamma_\text{int}} = \pfrac{\bm{d}}{t^\text{s}} \approx \bm{0}, \qquad \bm{x} \in \Gamma_\text{int},
\label{eq_int_velo}
\end{alignat}
which differs from the typical \gls*{fsi} condition that non-zero fluid and solid velocities must match at the interface. The interface condition \eqref{eq_int_velo} is equivalent to a no-slip condition in a rigid-wall fluid simulation. The traction interface condition
\begin{alignat}{3}
\bs{\sigma}^\text{s}_{\Gamma_\text{int}} \cdot \bm{n} = \bs{\sigma}^\text{f}_{\Gamma_\text{int}} \cdot \bm{n}, \qquad \bm{x} \in \Gamma_\text{int},
\end{alignat}
is identical to most \gls*{fsi} problems, with interface normal vector $\vc n$. While blood pressure is typically assumed to be constant in purely solid \gls*{gr} models, the \gls*{fsge} couples the normal components of local solid and fluid stress tensors, $\bs{\sigma}^\text{s}_{\Gamma_\text{int}}$ and $\bs{\sigma}^\text{f}_{\Gamma_\text{int}}$, respectively, which spatially vary according to the solution of the Navier-Stokes equations.

The choice of coupling scheme is crucial for the robustness and accuracy of the \gls*{fsge} model. For a comprehensive review of \gls*{fsi} coupling approaches, we refer to Hou~et~al. \cite{hou12}. We refer to Degroote \cite{degroote13} for an extensive discussion of partitioned fluid-structure coupling approaches and reuse some of the notations in this section. The \gls*{fsge} model proposed in this work is available open-source in the Python coupling code \texttt{svFSGe} \cite{svFSGe}. In the remainder of this section, we briefly review key differences between \textit{monolithic} (\ref{sec_monolithic}) and \textit{partitioned} (\ref{sec_partitioned}) coupling schemes for their use in \gls*{fsge}. The monolithic approach simultaneously solves the space- and time-discretized fluid and solid fields. In contrast, the partitioned approach solves fluid and solid fields separately and can be implemented with either weak (\ref{sec_partitioned_weak}) or strong (\ref{sec_partitioned_strong}) coupling.

\subsection{Monolithic coupling\label{sec_monolithic}}
Let $\vc{f}$ and $\vc{s}$ denote the fluid and solid residual, respectively. The vector $\vc U$ represents the discrete fluid variables velocity and pressure; the vector $\vc D$ represents the discrete solid displacements. The monolithic \gls*{fsi} problem is then given by
\begin{equation}
\begin{cases}
\begin{aligned}
\vc f (\vc U, \vc D) &= \vc 0 \\
\vc s (\vc U, \vc D) &= \vc 0
\end{aligned}
\end{cases}.
\end{equation}
This system is commonly solved with the Newton-Raphson method. The update for fluid and solid increments, $\Delta\vc U$ and $\Delta\vc D$, respectively in iteration $k$ at a load step $t$ becomes
\begin{equation}
\begin{bmatrix}
\partial_{\vc U} \, \vc f  & \partial_{\vc D} \, \vc f  \\
\partial_{\vc U} \, \vc s  & \partial_{\vc D} \, \vc s
\end{bmatrix}_{t,k}
\begin{bmatrix}
\Delta\vc U \\
\Delta\vc D
\end{bmatrix}_{t,k+1}
= -
\begin{bmatrix}
\vc f \\
\vc s
\end{bmatrix}_{t,k},
\label{eq_fsi_monolithic}
\end{equation}
where $\partial_{\vc U} \, \vc f$ denotes the partial derivative of $\vc f$ with respect to $\vc U$. All blocks in Equation~\eqref{eq_fsi_monolithic} are treated in the same mathematical framework and implemented in \texttt{svFSIplus}. For \gls*{fsi} problems, the advantages of monolithic coupling can be faster convergence and higher accuracy than the partitioned approach. However, it may require more computational resources and a more intrusive software implementation. Independent of these challenges, we found a monolithic coupling approach to be intractable for an \gls*{fsge} problem. Solving Equation~\eqref{eq_fsi_monolithic} requires solving both fluid and solid fields with the same time step size, typically $T_\text{FSI} \sim 1\,\text{ms}$ or below in cardiovascular \gls*{fsi} problems \cite{schussnig22}. However, as discussed in Section~\ref{sec_fsge}, solid \gls*{gr} takes place over a much longer time scale than blood flow dynamics. In fact, the \gls*{cmme} returns the mechanobiological equilibrated \gls*{gr} response after an \textit{infinitely} long \gls*{gr} period $s \to\infty$ at every load step $t$. This limit renders solving Equation~\ref{eq_fsi_monolithic} with a uniform fluid-solid time step size intractable, and we, therefore, chose a partitioned scheme for \gls*{fsge}.

\subsection{Partitioned coupling\label{sec_partitioned}}
In partitioned FSI, fluid field $\mathcal{F}$ and solid field $\mathcal{S}$ solve for interface traction $\vc u$ and interface displacements $\vc d$, respectively so that 
\begin{alignat}{3}
\vc{u} &= \mathcal{F}(\vc{d}),\\
\vc{d} &= \mathcal{S}(\vc{u}).
\end{alignat}
In the fluid field $\mathcal{F}$, in addition to solving blood flow dynamics (Section~\ref{sec_eqs_fluid}), we apply a deformation to the computational grid of $\Omega^f$ to avoid excessive fluid mesh distortion. This is similar to the arbitrary Lagrangian-Eulerian (ALE) formulation typically employed in \gls*{fsi} problems where an arbitrary grid velocity is applied. However, due to \eqref{eq_int_velo}, the mesh deformation is quasi-static in \gls*{fsge}. In the \gls*{fsge} model, we apply an interface mesh deformation $\vc{d}_\text{int}$ and solve a linear elasticity problem to solve for the fluid mesh deformation
\begin{equation}
\vc K \cdot \vc{d}_f = \vc 0, \quad \text{with~} \vc{d}_f = \vc{d}_\text{int} \text{~on~} \Gamma_\text{int},
\label{eq_fluid_mesh}
\end{equation}
with a stiffness matrix $\vc K$. Inspired by Degroote \cite{degroote13}, we outline the steps within the fluid and solid solvers in Algorithms~\ref{alg_solver_fluid} and~\ref{alg_solver_solid}, respectively. Partitioned \gls*{fsi} offers the advantage of reusing and combining existing fluid and solid solvers with solution techniques targeted to the respective domain. Note that while independent solvers could be used for $\mathcal{F}$ and $\mathcal{S}$, we use our open-source multi-physics solver \texttt{svFSIplus} for both \cite{zhu22b}. Crucially, this allows us to naturally split both time scales arising in the \gls*{fsge} problem. We thus focus on a partitioned coupling approach for the remainder of this work.
\begin{center}
\begin{minipage}{0.32\textwidth}
\begin{algorithm}[H]
\begin{algorithmic}[1]
\State Apply interface displacement $\vc d$
\State Deform the fluid domain \eqref{eq_fluid_mesh}
\State Solve for fluid variables $\vc U$
\State Extract interface traction $\vc u$
\end{algorithmic}
\caption{$\mathcal{F}$: Steady-state fluid \label{alg_solver_fluid}}
\end{algorithm}
\end{minipage}~
\hspace{1cm}
\begin{minipage}{0.45\textwidth}
\begin{algorithm}[H]
\begin{algorithmic}[1]
\State Apply interface traction $\vc u$
\State Propagate WSS \revc{stimulus} through solid domain \label{alg_solver_solid_wss}
\State Solve for solid variables $\vc D$
\State Extract interface displacement $\vc d$
\end{algorithmic}
\caption{$\mathcal{S}$: Quasi-static solid \label{alg_solver_solid}}
\end{algorithm}
\end{minipage}
\end{center}
Note that in comparison to a classic \gls*{fsi} problem, in \gls*{fsge} we additionally need to radially propagate the effects of the \gls*{wss} at the interface throughout the thickness of the blood vessel in the solid domain (line~\ref{alg_solver_solid_wss} in Algorithm~\ref{alg_solver_solid}) as discussed in Section~\ref{sec_eqs_solid}. In a cylindrical reference domain with a structured solid grid, this can be easily facilitated with a cylindrical coordinate system. A radial coordinate can be defined in patient-specific geometries, e.g., by solving a diffusion problem commonly used in cardiac geometries \cite{africa23}.

\subsubsection{Weak coupling\label{sec_partitioned_weak}}
The group of partitioned \gls*{fsi} coupling schemes can be further split into implicit (or \textit{strong}), semi-implicit, and explicit (or \textit{weak}) schemes \cite{fernandez11}. As Section~\ref{sec_intro} outlines, most previous FSG approaches used weak coupling. Algorithm~\ref{alg_coupling_weak} illustrates the most basic case of a serial staggered weak coupling scheme. Here, the fluid solution is updated once in every load step $t$ based on the solid solution $\vc{d}_{t-1}$ from the previous load step. The updated fluid solution is then input to update the solid solution $\vc{d}_t$. For \gls*{fsge}, this weak coupling method is unstable for $\gain>0$ and is thus not further pursued in this work. Although weakly coupled \gls*{fsi} problems can sometimes be stabilized by reducing the time step size, this strategy is not possible in \gls*{fsge}. While an applied stimulus can be spread over more load steps, the time-independent solid \gls*{cmme} model always yields a long-term evolved equilibrium at $s\to\infty$.
\begin{center}
\begin{minipage}{0.4\textwidth}
\begin{algorithm}[H]
\begin{algorithmic}[1]
\State $\vc{d}_0 = \vc{0},~t=0$
\Do
    \State $t++$
    \State $\vc{d}_{t} = \mathcal{S} \circ \mathcal{F}(\vc{d}_{t-1})$
\doWhile{$t \leq t_\text{max}$}
\end{algorithmic}
\caption{Weak coupling scheme \label{alg_coupling_weak}}
\end{algorithm}
\end{minipage}
\end{center}

\subsubsection{Strong coupling\label{sec_partitioned_strong}}
%
Strong coupling schemes enforce an equilibrium between fluid and solid fields in every load step, which requires an additional iteration loop. Strong coupling is essential to solve the \gls*{fsge} problem when there is a strong interaction between fluid and solid, which is the case for $\gain>0$, that is, when the solid \gls*{gr} material depends on local \gls*{wss}. We control the convergence towards an equilibrium in coupling iterations $k$ at every load step $t$ with the displacement residual
\begin{equation}
\vc{r}_{t,k} = \dtk - \dk,
\label{eq_coup_residual}
\end{equation}
In the following, $\dtk$ denotes the output of a subsequent fluid-solid solve
\begin{equation}
\dtk = \mathcal{S} \circ \mathcal{F}(\dk),
\end{equation}
in contrast to the input $\dk$. Equilibrium between fluid and solid is achieved when the relative displacement error norm
\begin{equation}
\frac{\Vert \vc{r}_{t,k} \Vert_2}{\Vert \dk \Vert_2} < \epsilon_0, 
\end{equation}
is smaller than a given tolerance $\epsilon_0$. \revb{Note that for a sufficiently small $\epsilon_0$, the converged solution of the strongly coupled partitioned scheme is identical to the solution of a monolithic scheme.} The most basic form of strong coupling is the Gau\ss-Seidel iteration scheme depicted in line~\ref{alg_fsge_step1} of Algorithm~\ref{alg_fsge}. Here, fluid mesh deformation, fluid, and solid are executed sequentially, and we use the output from the previous coupling iteration as the input for the current iteration:
\begin{equation}
\dk = \tilde{\vc{d}}_{t,k-1}.
\label{eq_gauss_seidel}
\end{equation}
Gau\ss-Seidel schemes typically converge slowly or might not converge at all \cite{degroote13}. A straightforward approach to stabilize the Gau\ss-Seidel scheme is to introduce static relaxation
\begin{equation}
\dk = \omega \, \tilde{\vc{d}}_{t,k-1} + (1-\omega) \, \vc{d}_{t,k-1}, \quad 0<\omega\leq 1,
\label{eq_damp_static}
\end{equation}
with a static relaxation parameter $\omega$, which reduces to the Gau\ss-Seidel scheme \eqref{eq_gauss_seidel} for $\omega=1$. While usually improving numerical stability for $0<\omega<1$, this coupling scheme typically converges slowly and highly depends on the choice of the relaxation parameter $\omega$. A much more efficient scheme is obtained by dynamically choosing $\omega_k$ in each iteration $k$ through Aitken relaxation \cite{kuettler08,kuettler09}:
\begin{equation}
\omega_k = - \omega_{k-1} \frac{\vc{r}_{k-1} \cdot \Delta\vc{r}_k}{\Vert \Delta\vc{r}_k\Vert_2}, \quad\text{with}~\Delta\vc{r}_k = \vc{r}_k - \vc{r}_{k-1}.
\end{equation}
While easy to implement, Aitken relaxation can use only information from one previous coupling iteration and is thus limited in its potential to introduce stabilization to the coupling scheme. The interface quasi-Newton with inverse Jacobian from a least-squares model (IQN-ILS) can use information from up to $q$ previous coupling iterations \cite{degroote09,degroote13}. It is based on applying the Newton-Raphson method to solve the equation $\vc{r}_t=\vc{0}$, which yields
\begin{equation}
\dk = \tilde{\vc{d}}_{t,k-1} - \left(\left.\pfrac{\vc{r}}{\vc{d}} \right|_{t,k-1}\right)^{-1} \vc{r}_{k-1}.
\end{equation}
However, the Jacobian matrix $\partial\vc{r}/\partial\vc{d}$ is not known explicitly, and an approximation with finite differences would require several additional solver evaluations. In IQN-ILS, the Jacobian is approximated from a linear combination of known previous displacements, computing the input displacements as
\begin{align}
\dk = \tilde{\vc{d}}_{t,k-1} + \vc{W}\mathbf{c}, \quad\text{with}~ \vc{W} =[\Delta \tilde{\vc{d}}_{k-1}, \Delta \tilde{\vc{d}}_{k-2}, \dots],\quad\Delta \tilde{\vc{d}}_{k-1} = \tilde{\vc{d}}_{k} - \tilde{\vc{d}}_{k-1},
\end{align}
with coefficient vector $\vc{c}$ which we obtain by solving the least squares problem 
\begin{align}
\min_{\mathbf{c}} \Vert \vc{V}\mathbf{c}+\vc{r}_k\Vert_2, \quad\text{with}~ \vc{V} =[\Delta\vc{r}_{k-1}, \Delta\vc{r}_{k-2}, \dots].
\end{align}
This minimization problem is solved using the QR-decomposition $\vc{Q}\vc{R}=\vc{V}$. The matrices $\vc{V}$ and $\vc{W}$ are updated in every coupling iteration $k$ and truncated to contain a maximum of $q$ entries. We reuse information from previous load steps in $\vc{V}$ and $\vc{W}$ but never calculate increments across load steps. Furthermore, filtering ensures that linearly dependent columns $i_\epsilon$ are excluded from $\vc{V}$ and $\vc{W}$ when $\vc{R}_{ii}<\epsilon_{QR}$. The IQN-ILS approach is summarized in Algorithm~\ref{alg_iqn_ils}.
\begin{center}
\begin{minipage}{0.5\textwidth}
\begin{algorithm}[H]
\begin{algorithmic}[1]
\State $\vc{V}=\emptyset,~\vc{W}=\emptyset,~t=0$
\Do
    \State $k=0$
    \State $\vc{d}_{t,0}$ from Algorithm~\ref{alg_predictor}
    \Do
        \State $k++$
        \If{$t=0$}
            \State $\dk = \tilde{\vc{d}}_{t,k-1}$ \label{alg_fsge_step1}
        \ElsIf{$(k = 1)$ or ($t=1$ and $k\le5$)}
            \State $\dk = \omega \, \tilde{\vc{d}}_{t,k-1} + (1-\omega) \, \vc{d}_{t,k-1}$ \label{alg_fsge_step_ini}
        \Else
            \State $\dk$ from Algorithm~\ref{alg_iqn_ils} \label{alg_fsge_step2}
        \EndIf
        \State $\dtk = \mathcal{S} \circ \mathcal{F}(\dk)$

        \State $\vc{r}_{t,k} = \dtk - \dk$
    \doWhile{$\Vert \vc{r}_{t,k} \Vert_2 \,/\, \Vert \dk \Vert_2 > \epsilon_0$}
    \State $t++$
\doWhile{$t \leq t_\text{max}$}
\end{algorithmic}
\caption{Strongly coupled \gls*{fsge}\label{alg_fsge}}
\end{algorithm}
\end{minipage}~
\hspace{1cm}
\begin{minipage}{0.35\textwidth}
\begin{algorithm}[H]
\begin{algorithmic}[1]
\If{$t=0$}
    \State $\vc{d}_{t,\,0} = \vc{d}_{0}$
\Else
    \State $\vc{d}_{t,\,0} = 2\vc{d}_{1} - \vc{d}_{0}$
\EndIf
\end{algorithmic}
\caption{Polynomial predictor \label{alg_predictor}}
\end{algorithm}
\begin{algorithm}[H]
\begin{algorithmic}[1]
\State $\vc{V} ~= (\vc{r}_{t,k} \,- \vc{r}_{t,k-1}, \vc{V})$
\State $\vc{W} = (\dtk - \tilde{\vc{d}}_{t,k-1}, \vc{W})$
\State $\vc{V} = \vc{V} \setminus \vc{V}_{i>q},~ \vc{W} = \vc{W} \setminus \vc{W}_{i>q}$
\Do
    \State $\vc{Q}\vc{R} = \vc{V}$
    \State $i_\epsilon = \arg_i \left( |\vc{R}_{ii}| < \epsilon_\text{QR} \right)$
    \State $\vc{V} = \vc{V} \setminus \vc{V}_{i_\epsilon},~ \vc{W} = \vc{W} \setminus \vc{W}_{i_\epsilon}$
\doWhile{$i_\epsilon \neq \emptyset$}
\State $\vc{c} = \vc{R}^{-1} \,\vc{b}$ with $\vc{b} = - \vc{R}^{-\text{T}} \, (\vc{V} \vc{r}_{t,k})$
\State $\dk = \tilde{\vc{d}}_{t,k-1} + \vc{W}\vc{c}$
\end{algorithmic}
\caption{IQN-ILS \label{alg_iqn_ils}}
\end{algorithm}
\end{minipage}
\end{center}
The strongly coupled \gls*{fsge} method is outlined in Algorithm~\ref{alg_fsge}. We use a Gau\ss-Seidel scheme exclusively at $t=0$ for the pre-loading stage in \gls*{cmme} \cite{latorre20} in line~\ref{alg_fsge_step1}. Here, the displacements are negligibly small, and there is only weak coupling as pre-loading does not depend on local \gls*{wss}. The \gls*{gr} stage in \gls*{cmme} for all $t>0$ requires the more robust IQN-ILS coupling approach in line~\ref{alg_fsge_step2}. Note that the performance of the algorithm also crucially depends on an accurate prediction of displacements $\vc{d}_{t,0}$ at the beginning of each load step. If a previous load step is available, we use the linear predictor in Algorithm~\ref{alg_predictor}. Further, note that at the beginning of each coupling iteration $k=1$, we obtain input displacements $\dk$ from a static relaxation step \eqref{eq_damp_static}. Similarly, we use static relaxation for the first five coupling iterations at $t=1$, i.e., the first \gls*{gr} load step, to collect an initial number of iterations for IQN-ILS.

\section{Illustrative computational examples \label{sec_results}}
\glsresetall
This section compares our \gls*{fsge} model to a purely solid \gls*{gr} model. As an illustrative example, we focus on the growth of an aortic aneurysm in a mouse. Aortic aneurysms have been studied in prior work using the \gls*{cmme} \cite{latorre20,latorre20b} and a prior fluid-solid-growth coupling method \cite{figueroa09}. Given the inherently large deformation in aortic aneurysms, they locally alter blood flow \cite{les10} and are thus a good example to demonstrate differences between solid \gls*{gr} and \gls*{fsge} and a good starting step for more complex \textit{in vivo} flow patterns in future research.

\begin{figure}[H]
\centering
\includegraphics[width=.55\linewidth]{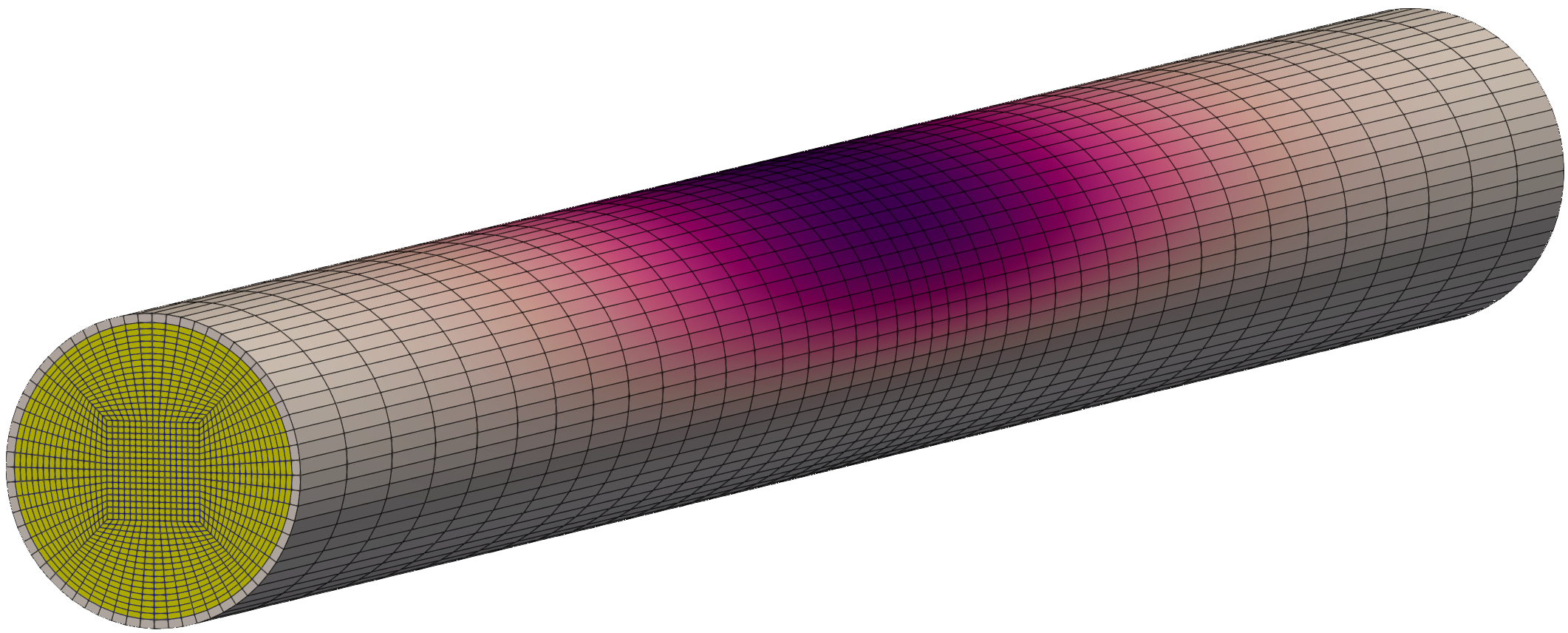}~
\includegraphics[width=.44\linewidth]{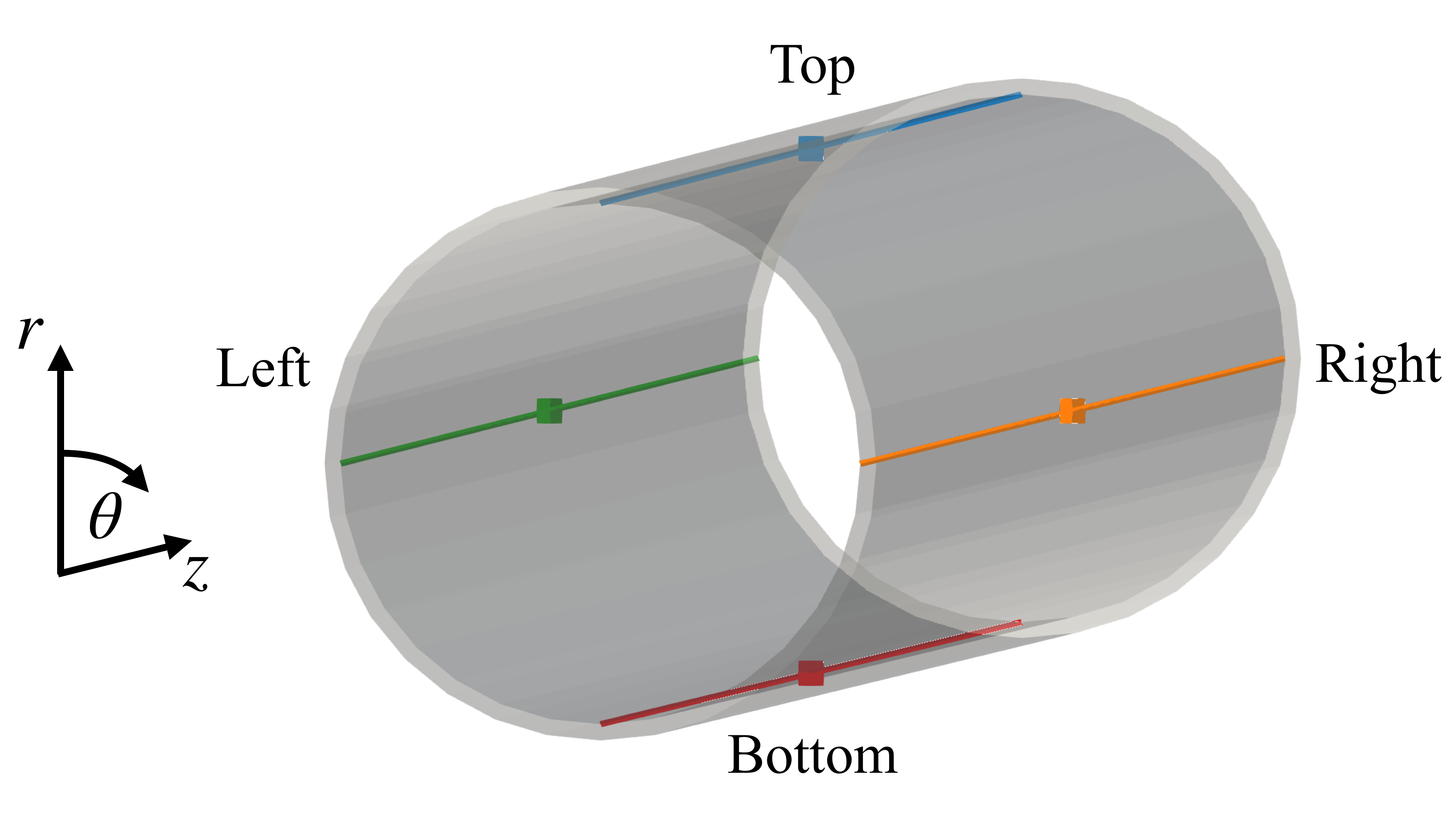}
\caption{Mesh (left) with fluid domain $\mathcal{F}$ (yellow) and solid domain $\mathcal{S}$, colored by the spatial insult profile $f_\theta \, f_z$ (white healthy, purple initiating insult profile). Plot locations (right), top, right, bottom, left, with a cylindrical coordinate system $(r,\theta,z)$.\label{fig_locations}}
\end{figure}

\begin{table}[htbp]
\renewcommand{\arraystretch}{1.1}
\footnotesize
\centering
\begin{tabular}{|lll|}
\hline
\textbf{Solid \gls*{gr} parameters \cite{latorre20}}&&\\
Constituents & $e,m,c$ & elastin, smooth muscle, collagen\\
Inner radius, thickness, length & $a_o, h_o, l_o$ & 0.647 mm, 0.04 mm, 15 mm\\
Mass fractions & $\phi^e_o, \phi^m_o, \phi^c_o$ & 0.34, 0.33, 0.33 \\
Collagen orientation fractions & $\beta_{\theta}, \beta_{z}, \beta_{d}$ & 0.056, 0.067, 0.877 \\
Diagonal collagen orientation & $\alpha_0$ & $29.9^\circ$ \\
Elastin stiffness & $c^e_o$ & 89.71 kPa \\
Smooth muscle parameters & $c^m_1,c^m_2$ & 261.4 kPa, 0.24 \\
Collagen parameters & $c^c_1,c^c_2$ & 234.9 kPa, 4.08 \\
Deposition stretches & $G^e_\theta, G^e_z, G^e_r, G^m, G^c$ & 1.90, 1.62, $1/(G^e_{\theta}G^e_{z})$, 1.20, 1.25 \\
Smooth muscle-to-collagen turnover ratio & $\eta$ & 1.0 \\
Shear-to-intramural gain ratio & $\gaino$ & \{0.0, 0.2, 0.4, 0.6, 0.8, 1.0\}\\
External tissue support stiffness & $k$ & 2.0 kPa/mm \\
\hline
\textbf{Aneurysm parameters}&&\\
Circumferential extent & $\theta_{od}$ & $0.55\,\pi$ \\
Circumferential decay & $v_\theta$ & 6 \\
Axial extent & $z_{od}$ & $l_o/4$ \\
Axial decay & $v_z$ & 2 \\
Maximum elastin degradation & $\varphi^e_{hm}$ & 0.7 \\
\hline
\textbf{Fluid parameters}&&\\
Maximum inflow velocity & $u_\text{in}$ & 1000 mm/s\\
Outlet pressure \cite{latorre20}& $p_\text{out}$ & 104.9 mmHg\\
Dynamic viscosity & $\mu$ & $4.0 \cdot 10^{-6}$ kg/mm/s \\
Density & $\rho$ & $1.06 \cdot 10^{-6}$ kg/mm$^3$ \\
\hline
\textbf{Numerical parameters}&&\\
Number of elements & $n_\theta \times n_r \times n_z$ & $64\times1\times40$\\
Number of \gls*{gr} load steps \cite{latorre20}& $t_\text{max}$ & 10 \\
Coupling tolerance & $\epsilon_0$ & $10^{-3}$ \\
Static relaxation parameter & $\omega$ & 0.1 \\
IQN-ILS filtering tolerance & $\epsilon_\text{IQN-ILS}$ & $10^{-1}$ \\
IQN-ILS number of old iterations & $q$ & 20 \\
\hline
\end{tabular}
\caption{Parameters used in all computational examples in this work.\label{tab_mat}}
\end{table}
For an overview of all parameters, see Table~\ref{tab_mat}, where the solid \gls*{gr} parameters are taken from Latorre and Humphrey \cite{latorre20}. All parameters in the \gls*{gr} model are identical to the \gls*{fsge} model, with the \gls*{fsge} having additional fluid and coupling parameters. The inflow velocity $u_\text{in}$ was chosen as an upper bound for systolic velocities observed in mouse aortas \cite{feintuch07}. Aneurysmal growth is associated with elastin degradation, modeled here by decreasing elastin stiffness \cite{watton04}, i.e., $c^e_h\to0$, and loss of mechanosensation, i.e., $\gainh\to0$ \cite{humphrey14}. We apply a spatial insult profile to trigger asymmetric aneurysmal growth,
\begin{align}
c^e_h(\theta,z,t) &= c^e_o \, [1 - \varphi^e_{hm} \, f(\theta,z,t)], \\
\gainh(\theta,z,t) &= \gaino \, [1 - f(\theta,z,t)], 
\end{align}
reducing long-term evolved elastin stiffness $c^e_h$ and gain ratio $\gainh$ with the spatial and temporal factor $f$ and maximum elastin degradation factor $\varphi^e_{hm}$. In our computational results, we sampled aneurysmal growth for different original homeostatic gain ratios $\gaino\in[0,1]$ as it proved crucial for the \gls*{gr} vs. \gls*{fsge} comparison. The circumferential, axial, and temporal factors are
\begin{equation}
\begin{aligned}
f(\theta,z,t) = f_\theta \, f_z \, f_t, \quad f_\theta = \exp \left( - \left| \frac{\theta_o - \pi}{\theta_{od}} \right|^{v_\theta} \right), \quad
f_z = \exp \left( - \left| \frac{z_o}{z_{od}} - \frac{1}{2} \right|^{v_z} \right), \quad
f_t = \frac{\text{tanh}\,(2t/t_\text{max})}{\text{tanh}\,(2)} , \quad t>0,
\end{aligned}
\end{equation}
where the insult is ramped up over the load steps $t_\text{max}$. Recall that load step $t=0$ corresponds to the \gls*{cmme} pre-loading stage, whereas $t>0$ corresponds to \gls*{gr}. We modulate the temporal factor $f_t$ with the hyperbolic tangent to yield approximately equal displacement increments in all load steps; this enables using a linear predictor that initializes each load step with a solution close to the converged one. The spatial insult profile $f_\theta \, f_z$ is visualized in Figure~\ref{fig_locations} (left). We introduce an aneurysmal region locally at the top of the blood vessel, whereas the ends and the bottom maintain healthy parameters.

We generated the cylindrical structured mesh in Figure~\ref{fig_locations} (left) with hexahedral linear displacement-based finite elements that are refined at the location of the aneurysm in the middle. Further, the mesh has a conforming interface between fluid and solid domains. The global $(\theta,r,z)$ coordinate system and four different plot locations within the blood vessel are shown in Figure~\ref{fig_locations} (right). All results in this work are extracted at the fluid-solid interface, i.e., at the endothelium. We show plots over the axial length of the vessel and point-wise results at the apex of the aneurysm.

\subsection{Computational performance}
\begin{figure}[H]
\centering
\includegraphics[width=.6\linewidth]{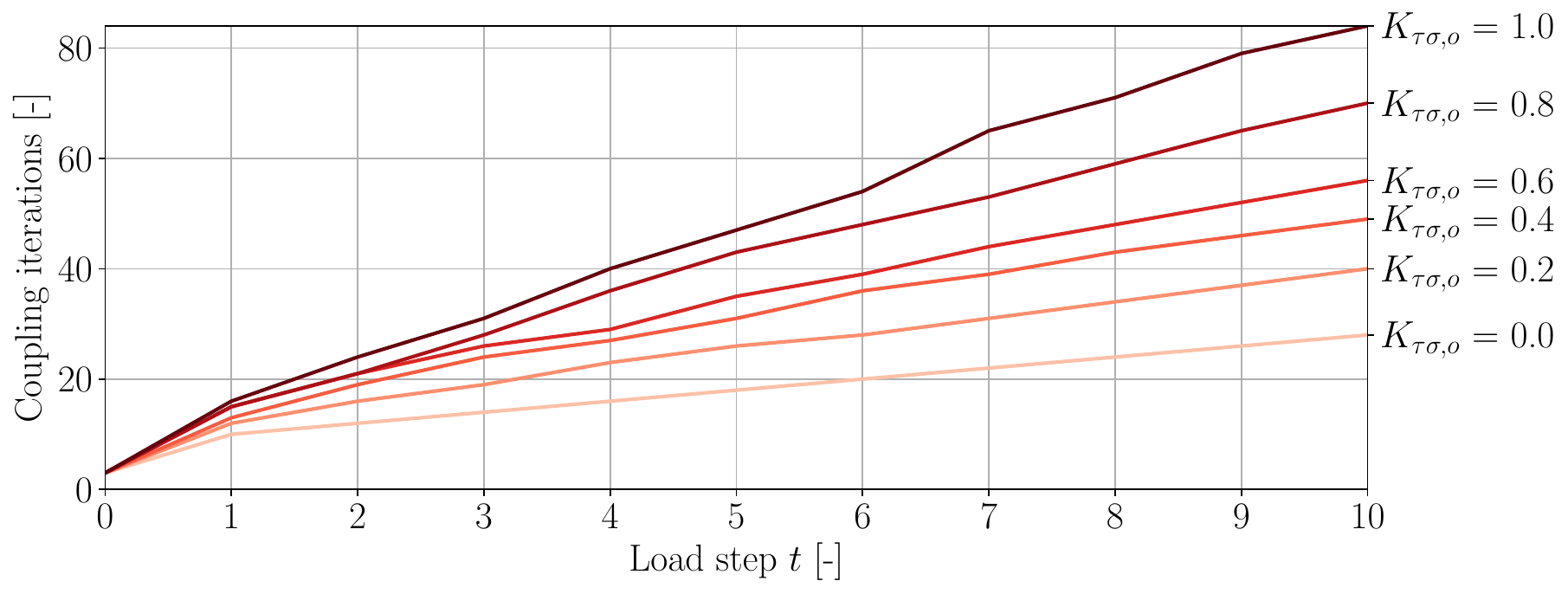}
\caption{Cumulative number of \gls*{fsge} coupling iterations per load step for different gain ratios $\gaino$.\label{fig_convergence}}
\end{figure}

Figure~\ref{fig_convergence} shows the cumulative number of \gls*{fsge} coupling iterations in each load step for varying gain ratios $\gain\in[0,1]$ required to reach the coupling tolerance $\epsilon_0$. Load step $t=0$ denotes the \gls*{cmme} pre-loading step, which takes only three iterations to converge in all simulations since the displacements $\vc d$ are close to zero during pre-loading. Our IQN-ILS coupling scheme from Algorithm~\ref{alg_iqn_ils} relies on solutions from previous coupling iterations, which are not yet available at the beginning of load step $t=1$. In addition, since there is only one prior converged load step, we cannot use extrapolation for the predictor. Thus, load step $t=1$ requires more coupling iterations than other load steps in all simulations. Importantly, there is a clear trend that more coupling iterations are required with higher $\gaino$. The average number of coupling iterations for load steps 2 to 10 ranges from 2.0 for $\gaino=0.0$ to 7.6 for $\gaino=1.0$. For the longest simulation $\gaino=1.0$ with 84 coupling iterations, the total wall times on an Apple M2 Max CPU were 2.7 min (one core) for linear-elastic mesh deformation and 52 min (six cores) for fluid dynamics in the fluid domain $\mathcal{F}$ and 1.7 min (one core) for \gls*{gr} with \gls*{cmme} in the solid domain $\mathcal{S}$. This renders the steady-state fluid simulation the computational bottleneck of \gls*{fsge}.

\subsection{Solid mechanics\label{sec_res_solid}}
\begin{figure}[H]
\centering
\includegraphics[width=\linewidth]{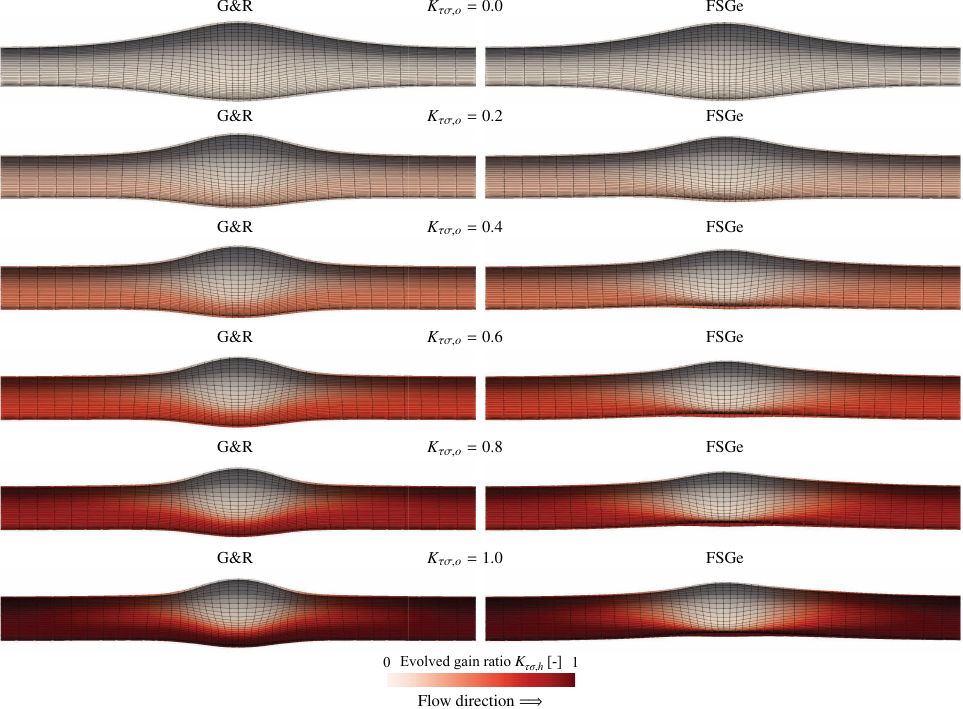}\\
\caption{Long-term evolved solid domains in \gls*{gr} (left) vs. \gls*{fsge} (right) for different original gain ratios $\gaino\in[0,1]$ (from top to bottom). The geometries are colored by the evolved gain ratio $\gainh$.\label{fig_meshes_kski_solid}}
\end{figure}

This section compares the solid mechanics between solid-only \gls*{gr} simulations and coupled \gls*{fsge} simulations, both using \gls*{cmme} as the solid model. Figure~\ref{fig_meshes_kski_solid} shows the displacement fields for \gls*{gr} (left column) and \gls*{fsge} (right column) for increasing original homeostatic gain ratios $\gaino\in[0,1]$ from top to bottom. The red color corresponds to the evolved gain ratio $\gainh$ after applying the aneurysm insult. Note that the evolved blood vessel retains its healthy original homeostatic parameters outside the top part of the aneurysm where the aneurysm insult was applied. Generally, a smaller original homeostatic gain ratio $\gaino$ leads to larger aneurysm growth. Comparing the top row with uniform $\gaino=0.0$, there is little difference between \gls*{gr} and \gls*{fsge} results. However, the aneurysms are smaller with \gls*{fsge} than with \gls*{gr} for $\gaino>0$. The differences become more pronounced for increasing gain ratio $\gaino\to1.0$, especially in the bottom aneurysmal region with intact elastin stiffness and mechanosensation. Interestingly, the radial deformation in this region is close to zero for $\gaino=0.2$ and becomes negative, i.e., inward, for $\gaino>0.2$.

\begin{figure}[H]
\footnotesize
\centering
\includegraphics[width=\linewidth]{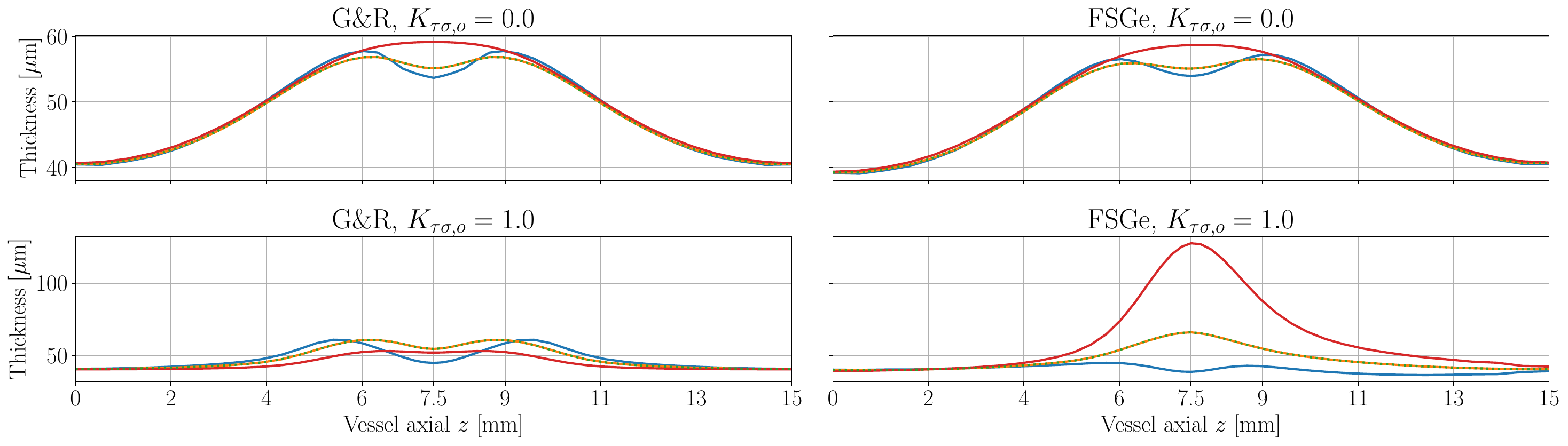}\\
Flow direction~$\Longrightarrow$
\caption{Thickness according to locations in Fig.~\ref{fig_locations}: top (blue), bottom (red), sides (orange/green). \label{fig_thick}}
\end{figure}

Figure~\ref{fig_thick} shows the local thickness over the axial length, $z$, of the blood vessel for $\gaino\in\{0.0,1.0\}$, i.e., for the top and bottom rows in Figure~\ref{fig_meshes_kski_solid}. Again, \gls*{gr} and \gls*{fsge} models were very similar for $\gaino=0.0$ (top row). While all \gls*{gr} results 
are symmetric, we observed minor thickening of about 3\% downstream at $z=11$ compared to upstream at $z=4$ for the \gls*{fsge} model. With $\gaino=1.0$, there was a pronounced axial asymmetry in the \gls*{fsge} model (bottom right) compared to the \gls*{gr} model (bottom left). Comparing the same locations, we observed 12\% downstream thinning at the top of the aneurysm (blue) and 24\% thickening at the bottom (red). Most importantly, the thickness at the bottom of the aneurysm (red) reaches an evolved equilibrial thickness of more than three times the original homeostatic value.

\begin{figure}[H]
\footnotesize
\centering
\includegraphics[width=\linewidth,trim={0 0 0 0},clip]{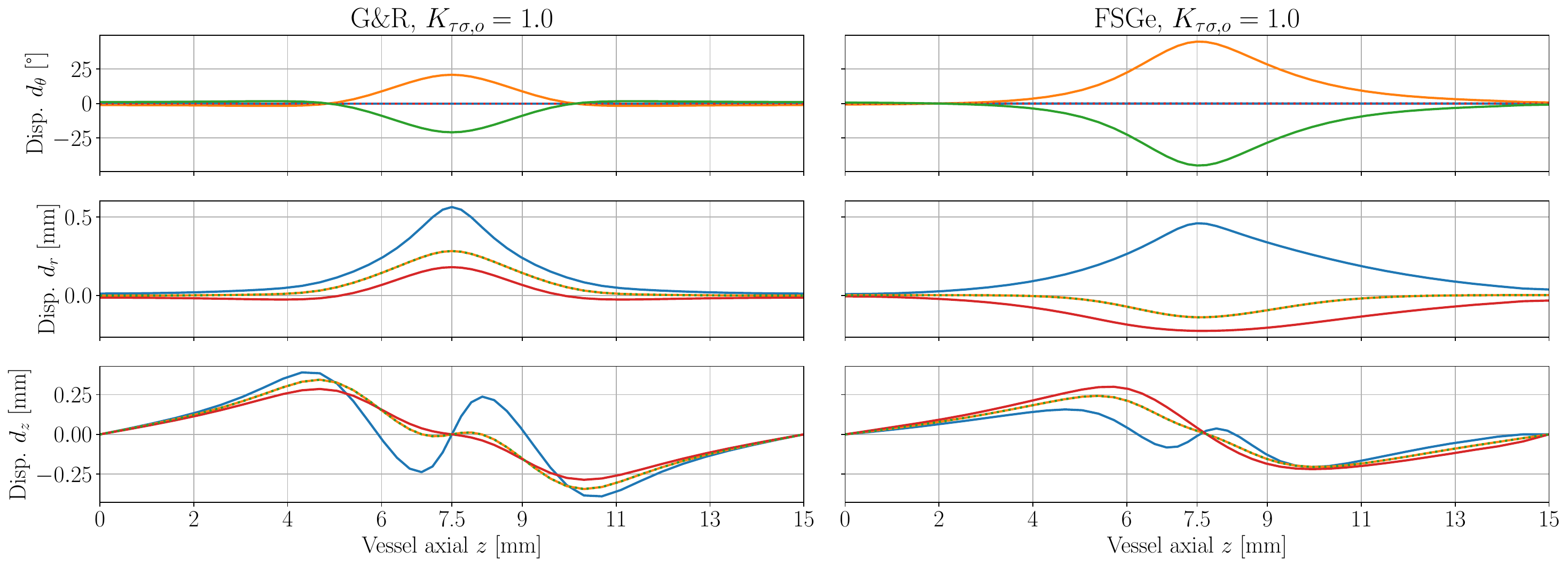}\\
Flow direction~$\Longrightarrow$
\caption{Displacements $d_\theta,d_r,d_z$ (top to bottom) at the endothelium according to locations in Fig.~\ref{fig_locations}: top (blue), right (orange), bottom (red), left (green) for varying original homeostatic gain ratios $\gaino$. \label{fig_disp_axi_in_t10}}
\end{figure}

The local displacements $d_{\{\theta,r,z\}}$ in the cylindrical coordinate system are shown in Figure~\ref{fig_disp_axi_in_t10}. The comparison is shown for the highest sampled gain ratio $\gaino=1.0$ (bottom row of Figure~\ref{fig_meshes_kski_solid}), which showed the most pronounced differences between the two models. Due to the symmetry of the aneurysm, there were no circumferential displacements (top row) at the top and bottom (blue/red dotted) of the aneurysm in both models. However, at the sides of the aneurysm (orange, green), there was double the circumferential displacement, i.e. rotation, towards the bottom with \gls*{fsge} than with \gls*{gr}. The radial displacements (middle row) at the bottom (red) and the sides (orange/green dotted) of the aneurysm became negative in the \gls*{fsge} model, indicating inward growth in these regions. Given the symmetry of the \gls*{gr} model, axial displacements (bottom row) are symmetric around $z=7.5$\,mm but asymmetric in the \gls*{fsge} model.

\begin{figure}[H]
\footnotesize
\centering
\includegraphics[width=\linewidth]{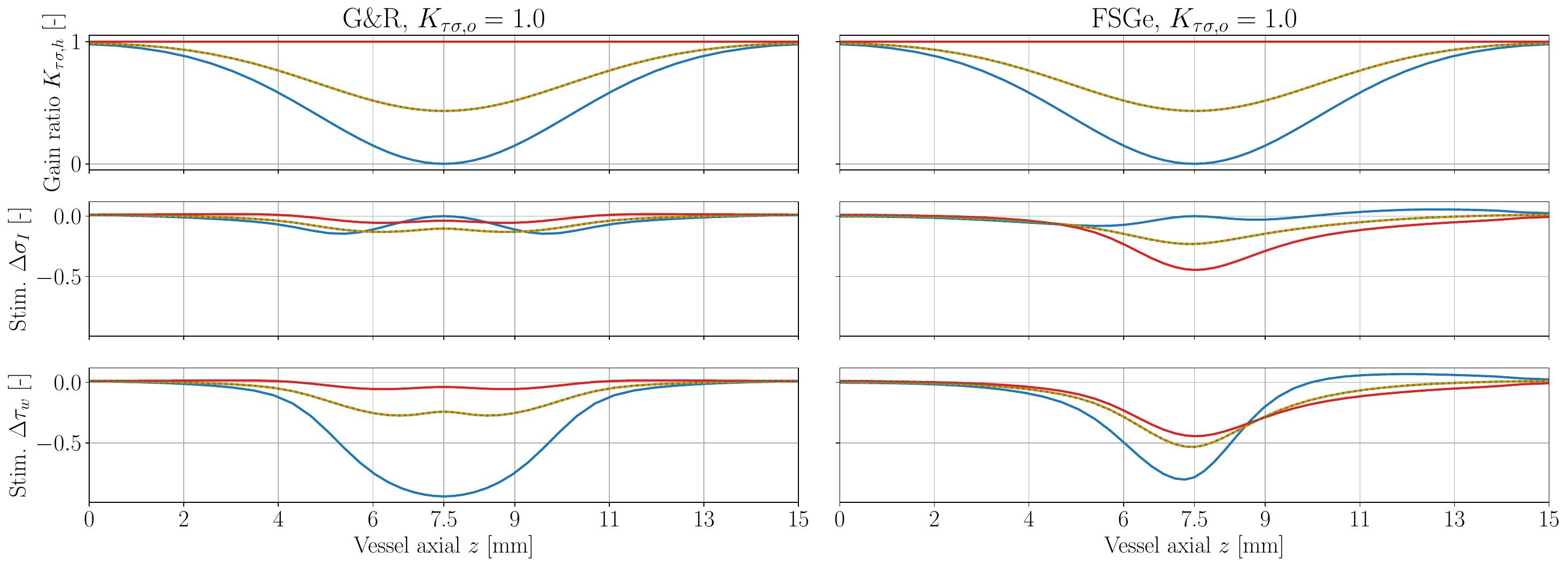}\\
Flow direction~$\Longrightarrow$
\caption{Prescribed evolved equilibrium gain ratio $\gainh=\stimi/\stimw$ (top) and resulting intramural $\stimi$ (middle) and \gls*{wss} stimuli $\stimw$ (bottom) in \gls*{gr} (left) and \gls*{fsge} (right) models. Locations according to Figure~\ref{fig_locations}: top (blue), bottom (red), sides (orange/green). \label{fig_stimuli}}
\end{figure}

The local mechanobiology in both \gls*{gr} and \gls*{fsge} for $\gaino=1.0$ is visualized in Figure~\ref{fig_stimuli}. The prescribed insult profile (top row) of mechanosensation loss towards the top of the aneurysm is identical in both models. However, the same prescribed gain ratios $\gainh=\stimi/\stimw$ are achieved with different levels of intramural stimuli $\stimi$ (middle row) and \gls*{wss} stimuli $\stimw$ (bottom row). Since $\gainh=0$ at the top peak of the aneurysm (blue curves at $z=7.5$), it follows for both models that $\stimi=0$ at this location. Furthermore, recall that we maintain intact mechanosensation at the bottom of the aneurysm (red curves at $z=7.5$). While in the \gls*{gr} model, the mechanobiologically equilibrated solution yields $\gainh=1.0\approx(-0.1)/(-0.1)$, the \gls*{fsge} model yields $\gainh=1.0\approx(-0.45)/(-0.45)$. This means that in the \gls*{gr} model's evolved equilibrium, both intramural stimulus $\stimi$ and \gls*{wss} stimulus $\stimw$ are reduced by 10\% compared to the original homeostatic state. However, in the \gls*{fsge} model, these were reduced by 45\%, thus arriving at a significantly different long-term evolved equilibrium that's characterized by the observed differences in local thickness (Figure~\ref{fig_thick}) and displacements (Figure~\ref{fig_disp_axi_in_t10}).

\begin{figure}[H]
\centering
\includegraphics[width=\linewidth]{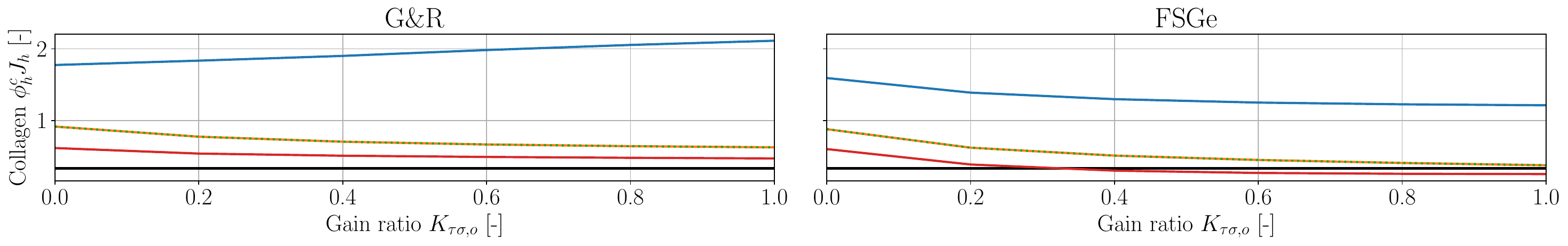}\\
\caption{Evolved referential collagen mass fraction (mass of collagen per unit reference volume) $\phi^c_h J_h$ at the endothelium at $z=7.5$\,mm for varying original gain ratios $\gaino$. The black line indicates the reference collagen mass of $\phi^c_o=0.33$. Locations according to Fig.~\ref{fig_locations}: top (blue), bottom (red), sides (orange/green). \label{fig_phic_curr_in_mid}}
\end{figure}

\begin{figure}[H]
\centering
\includegraphics[width=\linewidth,trim={0 0 0 0},clip]{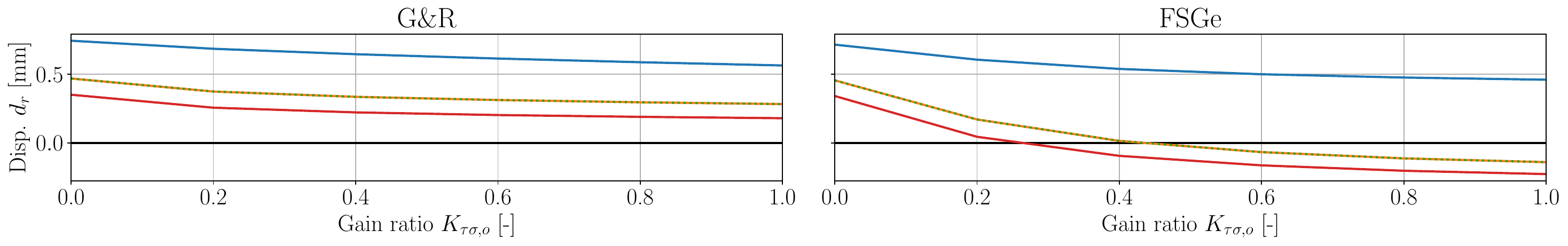}\\
\caption{Radial displacement $d_r$ at the endothelium at $z=7.5$\,mm for varying original gain ratios $\gaino$. Displacements $d_r<0$ (black line) indicate an inward growth of the aneurysm. Locations according to Fig.~\ref{fig_locations}: top (blue), bottom (red), sides (orange/green). \label{fig_disp_r_in_mid}}
\end{figure}

We visualize the evolved local collagen mass $\phi^c_h J_h$ in Figure~\ref{fig_phic_curr_in_mid} at the aneurysm mid-section over the sampled original gain ratios $\gaino$. Overall, collagen is deposited most at the top of the aneurysm (blue) and, to a lesser extent, at the healthy bottom (red). Notably, with increasing gain ratio $\gaino\to1$ in the \gls*{fsge} model (right), collagen is degraded at the bottom (red) below its initial mass fraction of $\phi^c_o=0.33$ (black). The degradation of collagen at the bottom (red) of the aneurysm is associated with inward growth around this region. This is visible in Figure~\ref{fig_disp_r_in_mid}, where radial displacements at the sides (green/orange) and bottom (red) become negative for $\gain\ge0.4$. These mechanisms are much less evident in the \gls*{gr} model, where there is little dependence on the original gain ratio $\gaino$ and no collagen degradation.

\subsection{Fluid dynamics\label{sec_res_fluid}}
\begin{figure}[H]
\centering
\includegraphics[width=\textwidth]{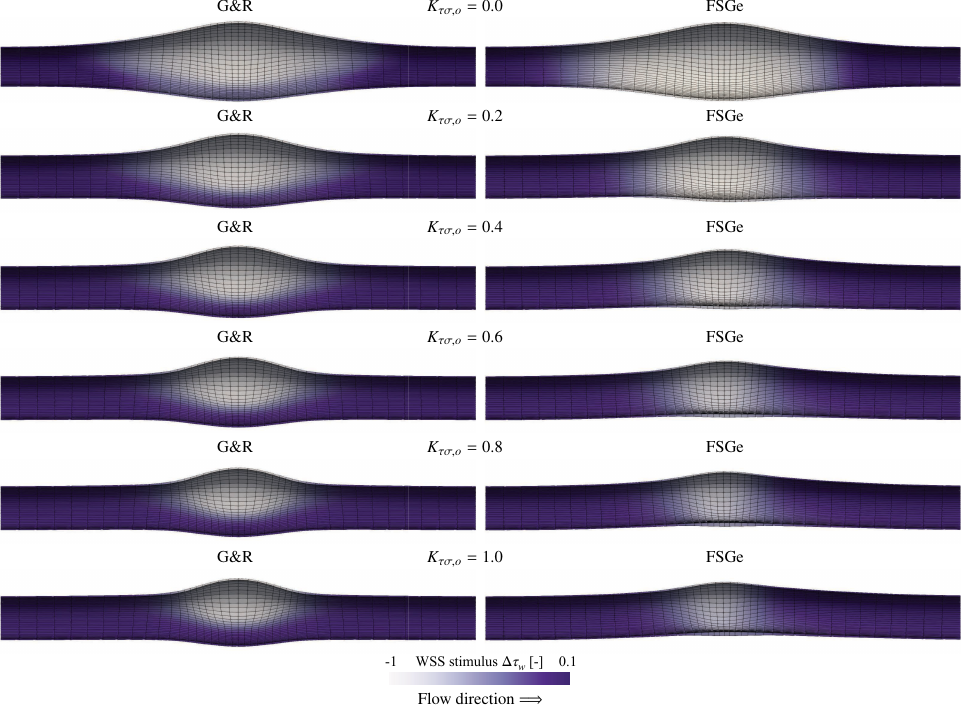}\\
\caption{\gls*{wss} stimulus $\stimw$ in \gls*{gr} (left) vs. \gls*{fsge} (right) for different original gain ratios $\gaino\in[0,1]$ (from top to bottom).\label{fig_meshes_kski_wss}}
\end{figure}

We compared solid mechanics in \gls*{gr} and \gls*{fsge} models in Section~\ref{sec_res_solid}. This section compares the fluid dynamics in both models. Figure~\ref{fig_meshes_kski_wss} shows the \gls*{wss} stimulus $\stimw$ for \gls*{gr} (left column) and \gls*{fsge} (right column) simulations. The plot locations are identical to Figure~\ref{fig_meshes_kski_solid}. Whereas the \gls*{wss} stimulus in the \gls*{fsge} model stems from solving the incompressible Navier-Stokes equations \eqref{eq_navier_stokes}, it is approximated from the Poiseuille solution \eqref{eq_poiseuille} in the \gls*{gr} model. Purple regions with a value of $\stimw\approx0$ correspond to regions that maintain \gls*{wss} during the development of the aneurysm. As expected, a higher gain ratio $\gaino\to1$ (from top to bottom) leads to more regions with $\stimw\to0$ due to the increasing weight of \gls*{wss} in finding the mechanobiological equilibrium. In \gls*{gr}, $\stimw$ is strictly coupled to the vessel's local change in endothelial radius, leading to strong circumferential variations in the apex of the aneurysm. In \gls*{fsge}, $\stimw$ is circumferentially more uniform around the deformed centerline of the blood vessel.

\begin{figure}[H]
\centering
\includegraphics[width=\linewidth]{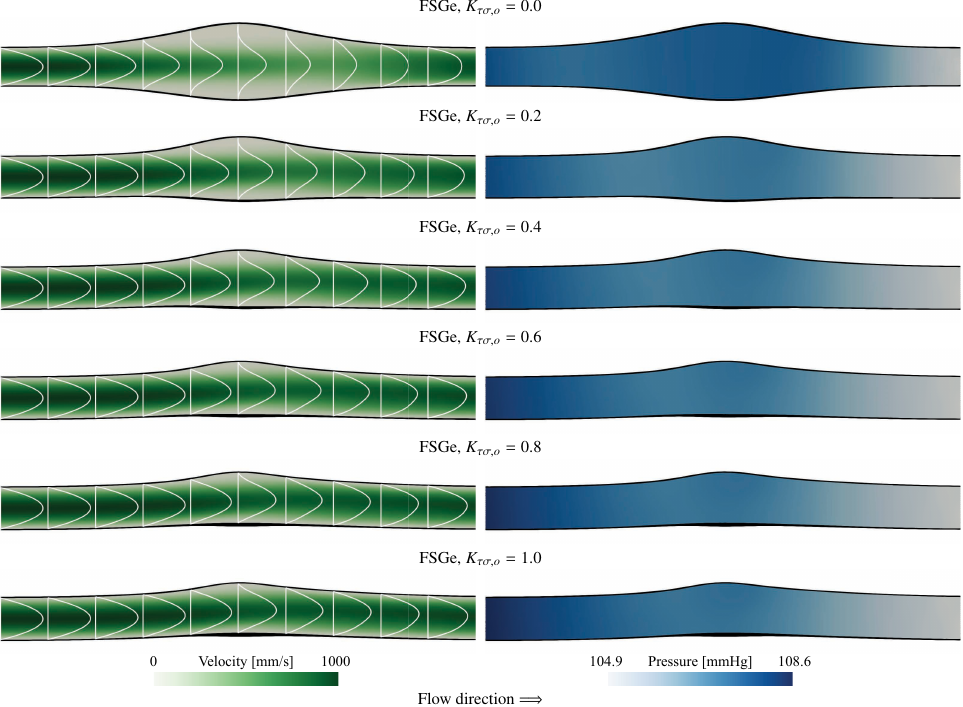}\\
\caption{Fluid domains in \gls*{fsge} models with velocity and flow profiles (left) and pressure (right) with vessel wall (black) for original homeostatic gain ratios $\gaino\in[0,1]$ (from top to bottom).\label{fig_meshes_kski_fluid}}
\end{figure}

Figure~\ref{fig_meshes_kski_fluid} shows velocity (left) and pressure (right) for the vessel's longitudinal midsection for different gain ratios $\gaino$ in the \gls*{fsge}. Due to the absence of local fluid dynamics in the \gls*{gr} model, we cannot compare the models. The geometries correspond to the right column in Figure~\ref{fig_meshes_kski_solid}. Recalling the fluid boundary conditions from Section~\ref{sec_results}, it is evident that all flows have identical (parabolic) inflow conditions on the very left and identical outflow pressures of 104.9\,mmHg on the very right. While there is little variation in pressure fields for the sampled gain ratios $\gaino$, there are some notable differences in the velocity field.  There is a small recirculation zone visible at the top apex of the aneurysm. Due to the incompressibility of blood, for $\gaino=0.0$ (top), the doubling in radius inside the aneurysm drastically decreases local velocity. This effect is less pronounced as the maximum radius decreases with increasing gain ratio $\gaino$. Interestingly, as the gain ratio approaches $\gaino\to1$, the velocity is deflected upward into the top part upstream of the aneurysm.

\begin{figure}[H]
\centering
\includegraphics[width=.49\linewidth]{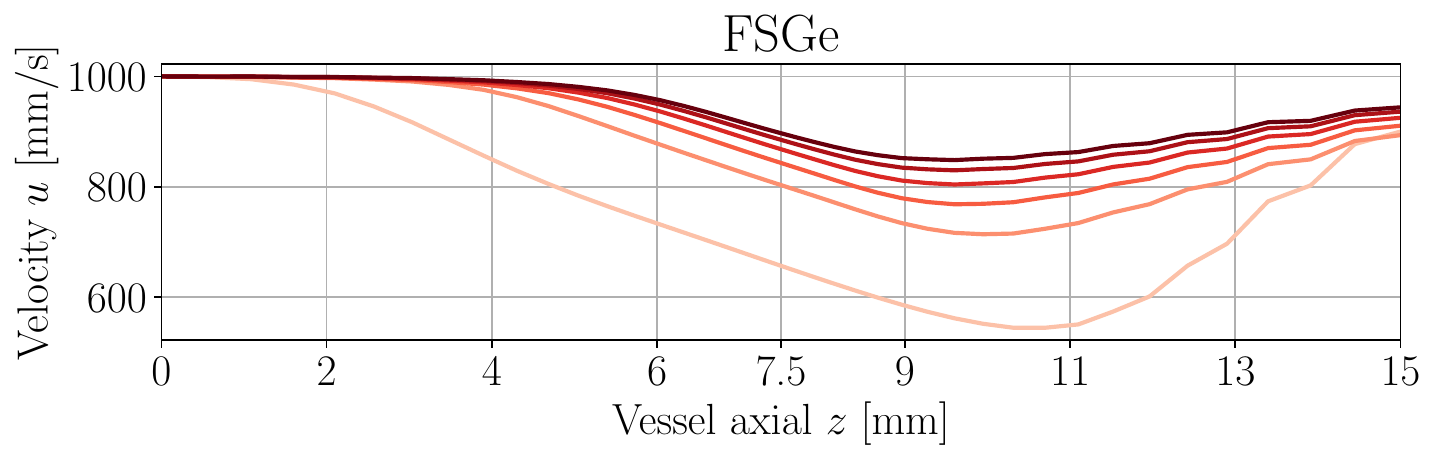}~
\includegraphics[width=.49\linewidth]{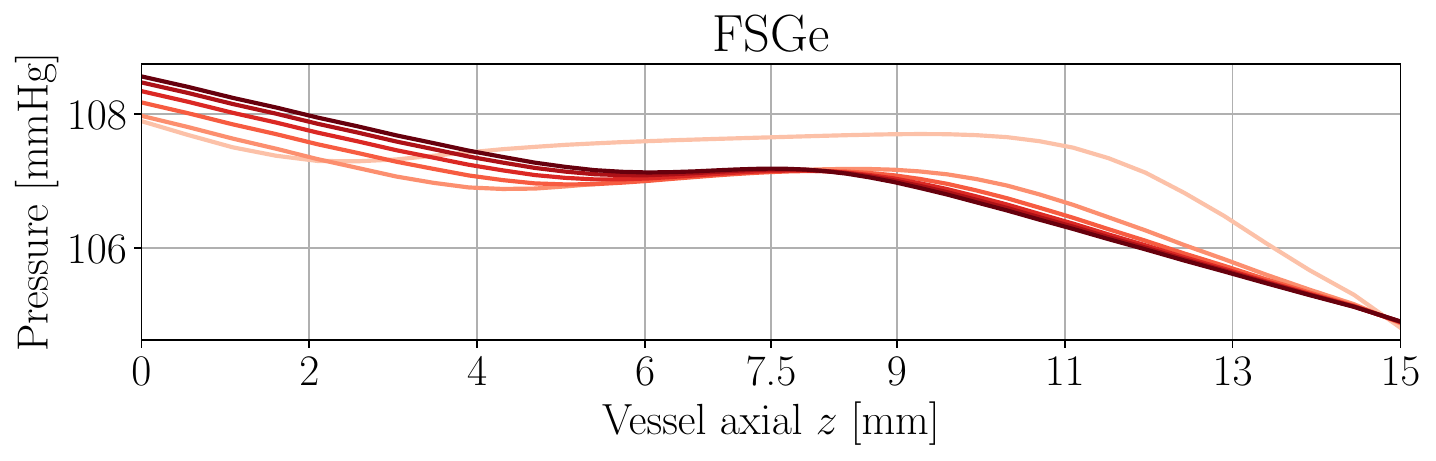}~
\caption{Velocity magnitude (left) and pressure (right) along the deformed vessel centerline from $\gaino=0.0$ (light) to $\gaino=1.0$ (dark). \label{fig_center_velocity_pressure}}
\end{figure}

 Figure~\ref{fig_center_velocity_pressure} shows fluid velocity $\bm{u}$ (left) and pressure $p$ (right) along the deformed centerline of the vessel for all gain ratios in Figure\ref{fig_meshes_kski_fluid}. Again, prescribed boundary conditions are evident for the inlet velocity at $z=0$\,mm and outlet pressure at $z=15$\,mm. The centerline velocity also coincides with the region of maximum velocity. The drop in velocity inside the aneurysm and the downstream acceleration distal to the aneurysm are visible. We also observe a linear drop in pressure along $z$ upstream and downstream of the aneurysm and a pressure plateau in the aneurysmal region. While there's a more than 60\% fluctuation in centerline velocity throughout all $\gaino$ samples, the pressure changes only about 3.5\% compared to the overall mean pressure.

\section{Discussion\label{sec_discussion}}
\glsresetall
We presented a novel \gls*{fsge} method, where we strongly couple the \gls*{3d} Navier-Stokes equations and the \gls*{3d} \gls*{cmme}. We demonstrated crucial differences in the solution compared to a solid-only \gls*{gr} model in illustrative computational examples of an asymmetric aortic aneurysm. We identified the original homeostatic gain ratio $\gaino$ as a crucial factor determining the degree of interaction between fluid and solid growth in the long-term evolution. In the remainder of this section, we discuss the results from Section~\ref{sec_results}, limitations, and future perspectives. We first review computational aspects of the \gls*{fsge} method proposed in this work: its equilibrated nature (\ref{sec_disc_transient}), its strong coupling (\ref{sec_disc_coupling}), stability of \gls*{cmme} (\ref{sec_disc_stability}), and original homeostasis (\ref{sec_disc_prestress}). We further review mechanobiological aspects of \gls*{fsge}, namely \gls*{wss} influence (\ref{sec_disc_wss}) and the directionality and pulsatility of blood flow (\ref{sec_disc_pulsatility}).

\subsection{Equilibrated vs. transient growth and remodeling\label{sec_disc_transient}}
Latorre and Humphrey \cite{latorre20} introduced the \gls*{cmme} as a fast, rate-independent version of the \gls*{cmm}. It directly yields the long-term equilibrated \gls*{gr} response, i.e., at $s\to\infty$. The computational cost of \gls*{cmme} is comparable to a solid mechanics FEM solution with a hyperelastic material. Together with our implicit coupling scheme, this makes the \gls*{fsge} highly efficient. On the other hand, it may be critical to model transient \gls*{gr} in cases where the insult is applied slowly compared to the \gls*{gr} response. For example, tissue-engineered vascular grafts in Fontan patients can develop a severe stenosis, which sometimes spontaneously resolves \cite{schwarz21,blum22}. Here, the time-resolved development of \gls*{gr} is crucial to studying the mechanisms underlying the stenosis formation and resolution \cite{schwarz23b}, which can be modeled with \gls*{fsge} \cite{latorre22}. However, we can only study phenomena that will yield a long-term mechanobiological equilibrium. Yet, many clinically interesting applications involve \gls*{gr} processes that do not necessarily yield an evolved equilibrial state \cite{ramachandra15}. For example, theories have been developed to quantify the stability or instability of \gls*{gr} processes such as aneurysm formation \cite{cyron14,latorre19}. Thus, transient \gls*{gr} models like the one developed by Schwarz et al. \cite{schwarz23b} will be necessary to study \gls*{fsg} in unstable cases.

\subsection{Strong vs. weak fluid-solid-growth coupling\label{sec_disc_coupling}}
We proposed a strongly coupled partitioned fluid-structure interaction model where the solid \gls*{gr} model is given by the \gls*{cmme}. A fully implicit monolithic coupling scheme is intractable in this case due to the discrepancy in time scales, $T_f \sim 1\,\text{ms}$ for the fluid and $T_s \to\infty$ for the solid \gls*{cmme}. Our coupling algorithm converged in only two coupling iterations for $\gaino=0.0$, allowing for a weakly coupled partitioned scheme. However, the number of coupling iterations rose with $\gaino\to1$, highlighting the strong interaction between fluid and solid. We experimented with static relaxation for \gls*{fsge} coupling. However, this required a very low relaxation parameter $\omega$ to stabilize the coupling scheme, which led to excessive coupling iterations. We achieved considerable speedups by introducing dynamic Aitken relaxation \cite{kuettler08} but could not find a solution for high $\gaino\to1$. Thus, IQN-ILS remains our coupling algorithm of choice \cite{degroote13}. More speedup could be achieved by taking into account the iterations internal to each solver \cite{spenke23}.

\subsection{Computational stability of the equilibrated constrained mixture model\label{sec_disc_stability}}
In addition to potential instabilities arising from the fluid-structure coupling in \gls*{fsge} (Section~\ref{sec_disc_coupling}), we discovered instabilities in the underlying solid \gls*{cmme} model for $\gaino\to1$, leading to oscillations in displacements, Jacobian, and internal \gls*{gr} variables. These are triggered by the growth of the aneurysm and can be interpreted as a mechanical instability problem. There may exist other biological equilibria of $\stimw$ and $\stimi$ that satisfy a certain gain ratio $\gaino>0$. Only the special case $\gaino=0=\stimi$ has a unique biological solution for $\stimi$. A mechanobiologically equilibrated solution is obtained by simultaneously solving for a mechanical equilibrium \eqref{eq_solid}. We believe these oscillations originate from evaluating the mechanobiological equilibrium \eqref{eq_stimulus_eq} point-wise on each Gau\ss-point, without having any continuity conditions for internal \gls*{gr} variables. Thus, it is possible that Gau\ss-points within the same element may converge to different mechanobiological equilibria. We currently prevent these oscillations by choosing only one element across the vessel thickness. A more versatile solution could be using an internal \gls*{gr} variable, e.g., collagen mass fraction $\phi^c_h$, as an additional unknown in addition to the displacements and choosing a lower polynomial order for its shape function. Element formulations could be adopted from (quasi)-incompressibility formulations, e.g., mixed-field approaches such as Q1P0 \cite{gueltekin18} or MINI elements \cite{karabelas19}. Alternatively, a functional could be added to the solid weak form preventing oscillations in collagen mass fraction. This idea is derived from edge stabilization in \gls*{cfd}, where a least-squares term penalizes the jump in the gradient of the discrete solution over element boundaries \cite{burman04}.

\subsection{Definition of original homeostasis\label{sec_disc_prestress}}
In computational biomechanics, the solid reference configuration, often obtained from \textit{in vivo} medical imaging, is typically not stress-free due to fluid or other loads. This requires either finding a stress-free reference configuration, which might not exist \cite{gee09}, or local residual stresses so that the initial configuration is in equilibrium with fluid loads. Many strategies have been developed to address this issue \cite{rausch17,genet19,barnafi24}. In \gls*{gr}, we also need to satisfy a \textit{biological} equilibrium, i.e., original homeostasis $o$, in addition to the aforementioned \textit{mechanical} equilibrium. In this work, we achieve a mechanobiological equilibrated original homeostasis by solving a nonlinear equation for a thin-walled cylinder for the initial \textit{in vivo} blood pressure $P_o$. Here, the inner and outer radii are fixed, as well as the homeostatic parameters in Table~\ref{tab_mat}. This approach works well in a given ideal cylinder with an unknown uniform \textit{in vivo} blood pressure that can be freely chosen in the model. However, it cannot be generalized to \textit{in vivo} blood vessel geometries that were imaged under measured blood pressure. Given the limited prior work with 3D \gls*{cmm}, there are few references for pre-loading \textit{in vivo} \gls*{gr} models. The algorithm proposed by Wu et al. \cite{wu20} solves a minimization problem by locally varying fiber alignment. Laubrie et al. \cite{laubrie22} prescribed a transmurally varying prestretch gradient for constituents to establish an initial homeostatic state. Gebauer et al. \cite{gebauer23} introduced an iterative algorithm to prestress a fixed patient-specific cardiac geometry under fixed ventricular pressures. The algorithm solves for the local isotropic elastin prestretch until the model is in mechanobiological equilibrium up to a specified displacement tolerance. The choice of elastin prestretch was motivated by the fact that elastin is the only constituent that is assumed to not turn over in the \gls*{gr} time scale considered, given its long half-life \cite{cocciolone18}. In the future, this approach could be used to define the original homeostasis in \gls*{fsge} problems by either locally solving for the elastin prestretch or other quantities, such as vessel thickness.

\subsection{Influence of local wall shear stress\label{sec_disc_wss}}
Previous \gls*{fsg} studies did not investigate differences across \gls*{gr} models or found only minor differences \cite{schwarz23b}. We observed in Section~\ref{sec_res_solid} that both \gls*{gr} and \gls*{fsge} models delivered similar results for $\gaino\to0$ despite the underlying assumptions in the purely solid \gls*{gr} model. However, for high gain ratios $\gaino\to1$, the different long-term evolved equilibria between \gls*{gr} and \gls*{fsge} models became more pronounced. These differences were especially prevalent in the bottom healthy region of the blood vessel with intact elastin stiffness and mechanosensation. Compared to \gls*{gr}, the \gls*{fsge} model grew \textit{inward} with a three-fold \textit{increase} in vessel wall thickness. Furthermore, in Section~\ref{sec_res_fluid}, we observed that the inward growth of the bottom region of the blood vessel is associated with an upward deflection of blood flow toward the top of the aneurysm.

As mentioned previously, earlier versions of \gls*{gr} relied on the Poiseuille and thin-walled cylinder assumptions to calculate mechanical stimuli. Assuming Poiseuille flow \eqref{eq_poiseuille} and a thin-walled cylinder for the blood vessel, analytic relationships can be derived for \gls*{wss} $\tau_w$ and wall stress $\sigma$ components, respectively,
\begin{align}
\tau_w \sim Q \cdot \frac{1}{\revc{a}^3}, \quad
\sigma_\theta \sim P \cdot \frac{\revc{a}}{h}, \quad
\sigma_r \sim -P, \quad
\sigma_z \sim F \cdot \frac{1}{\revc{a}h},
\end{align}
with \revc{luminal radius $a$}, wall thickness $h$, flow rate $Q$, pressure $P$, and axial force $F$. We specifically developed our \gls*{fsge} coupling method to \textit{not} rely on Poiseuille flow and thin-walled assumptions but rather the numerical solution of the \gls*{3d} Navier-Stokes equations and \gls*{3d} finite strain elasticity, respectively. Yet, we studied aneurysmal development under steady-state flow with a low Reynolds number of $\text{Re}\approx170$ which resulted in a flow profile similar to parabolic even in aneurysmal regions of the blood vessel. Thus, comparing the fluid and solid field solutions derived from our \gls*{fsge} model to these simplifying assumptions gives us insights into the feedback loop of biomechanically regulated variables (Figure~\ref{fig_fsg}) and helps us explain the differences in mechanobiological equilibria. In our examples, the flow rate $Q$ was constant, and there was little change in local pressure $P$ throughout the fluid domain. We can thus derive the following relationships for the change in \gls*{wss} and intramural stimuli
\begin{align}
\stimw &\sim \frac{\revc{a}_o^3}{\revc{a}_h^3} - 1,\label{eq_simplified_wss}\\
\stimi &\sim \frac{h_o}{\revc{a}_o}\frac{\revc{a}_h}{h_h} - 1.\label{eq_simplified_sigma}
\end{align}
The differences in aneurysmal growth were associated with these different definitions of stimuli and, therefore, different values of mechanobiological equilibria in the two models. Our illustrative computational examples highlight the importance of incorporating local fluid-derived \gls*{wss} in the \gls*{fsge} model. In the \gls*{fsge} model, the fluid couples the deformation in the aneurysmal top part of the blood vessel with the healthy bottom part. In the purely solid \gls*{gr} model, the \gls*{wss} in these regions is evaluated independently from local deformations. The stimuli $\stimw<0$ and $\stimi<0$ were downregulated to a much larger extent in \gls*{fsge} than in \gls*{gr} at the bottom of the aneurysm. In the \gls*{gr} model, relying on relationship \eqref{eq_simplified_wss} resulted in an increase in evolved radius $\revc{a}_h$ and thickness $h_h$. In the \gls*{fsge} model, the fluid-derived \gls*{wss}, $\stimw$ decreased with decreasing $\revc{a}_h$ which is contrary to \eqref{eq_simplified_wss}. This was possible due to the Navier-Stokes-derived \gls*{wss} (Figure~\ref{fig_meshes_kski_wss}, right) and the upward deflection of the blood flow inside the aneurysm (Figure~\ref{fig_meshes_kski_fluid}). In both models, this change was accompanied by an increase in thickness $h_h$ according to \eqref{eq_simplified_sigma}.

\subsection{Pulsatility and directionality of blood flow\label{sec_disc_pulsatility}}
Our \gls*{fsge} model considered steady-state flow conditions as an input for the intramural and \gls*{wss} \gls*{gr} stimuli, $\stimi$ and $\stimw$, respectively. This is equivalent to some previous \gls*{fsg} methods in Table~\ref{tab_fsg}, whereas others used time-averaged \gls*{gr} stimuli from a pulsatile flow simulation. Pulsatility is an important determinant of the blood vessel wall geometry, structure, and properties \cite{eberth10,kalenik20}. For example, Eberth et al. \cite{eberth09} found that increased wall thickness in hypertension correlated more strongly with pulse than with systolic and especially mean blood pressure. Extending our \gls*{fsge} approach to incorporate pulsatility would require solving an \gls*{fsi} problem with an elastic wall in each coupling iteration. We could then extract the instantaneous elastic properties of the \gls*{cmme} with a fixed \gls*{gr} state from the evolved equilibrial state, as shown by Latorre and Humphrey \cite{latorre20}. In the aneurysmal region, we currently introduce dysfunctional mechanosensing at the endothelium by regionally setting $\gaino\to0$ in the top part of the aneurysm \cite{humphrey14} as a model input. In the future, mechanosensing could be a function of pulsatility and thus as part of the model prediction.

\begin{figure}[H]
\footnotesize
\centering
\includegraphics[width=.7\linewidth,trim={0 36cm 0 0},clip]{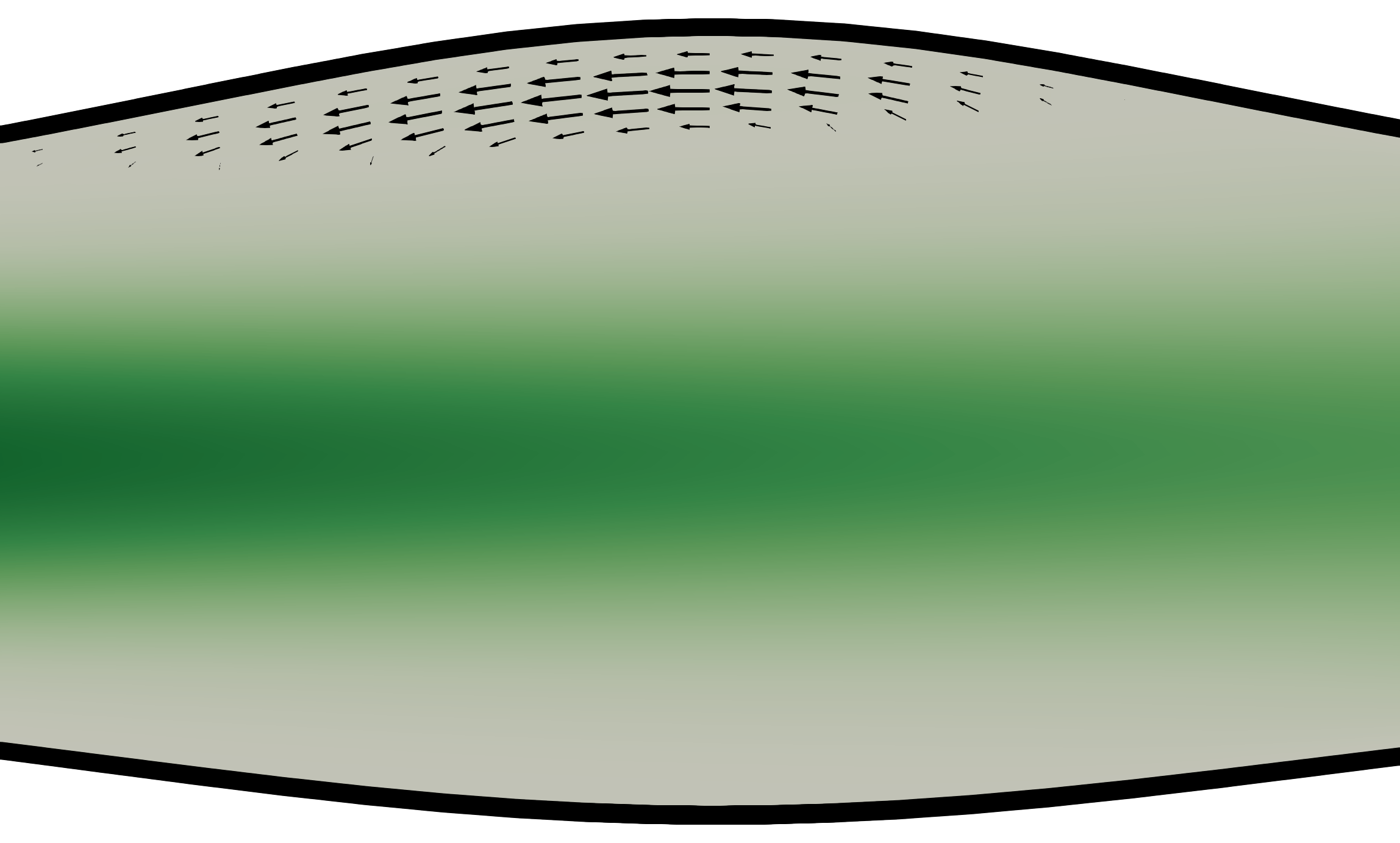}\\
Flow direction~$\Longrightarrow$
\caption{Recirculation zone in an \gls*{fsge} simulation with $\gaino=0.0$ (top left in Figure~\ref{fig_meshes_kski_fluid}), with backflow velocity vectors for $u_z<0$ at the top of the aneurysm. \label{fig_recirculation}}
\end{figure}

Another crucial input would be the variability of \gls*{wss} over a cardiac cycle. Figure~\ref{fig_recirculation} shows a recirculation zone with backflow against the predominant flow direction at the top of the aneurysm. This flow separation only occurs in the \gls*{fsge} model with $\gaino=0.0$ (top left in Figure~\ref{fig_meshes_kski_fluid}) due to its nearly two-fold increase in radius. Note, however, that the maximum backflow velocity is only 1.6\% of the maximum inlet velocity. This effectively reverses the direction of the local \gls*{wss} vector $\bs\tau_w$ acting on the vessel wall. However, in solid \gls*{gr} models, $\tau_w$ can only be estimated as a scalar quantity to inform the \gls*{wss} stimulus, typically derived from Poiseuille flow \eqref{eq_poiseuille}. It is well established that not only the magnitude but also the pulsatility and direction of \gls*{wss} are constantly sensed by the endothelium \cite{davies08,wang13}. For example, the directionality of \gls*{wss} plays an important role in the development of atherosclerosis \cite{baeyens16}. \reva{A directional metric of \gls*{wss} has been incorporated in the previous \gls*{fsg} approach of Teixeria et al. \cite{teixeira20}. Here, the oscillatory nature of flow is quantified by the wall shear stress aspect ratio, which captures the multi-directionality of flow \cite{vamsikrishna20}. In addition,} Mohamied et al. \cite{mohamied14} proposed a new metric,
\begin{equation}
\text{transWSS} = \frac{1}{T} \int_0^T \frac{\bs\tau_w \cdot \left(\bm n \times \bs\tau_\text{mean} \right)}{\Vert\bs\tau_\text{mean}\Vert} \,\text{d}t, \quad \bs\tau_\text{mean} = \frac{1}{T} \int_0^T \bs\tau_w \,\text{d}t,
\label{eq_transwss}
\end{equation}
to quantify the change of direction of \gls*{wss} over the course of a cardiac cycle $T$, with vessel normal vector $\bm{n}$ and mean \gls*{wss} vector $\bs\tau_\text{mean}$. This metric was shown to correlate better with atherosclerotic lesion prevalence than other metrics like time-averaged or oscillatory \gls*{wss} index. \revb{Our \gls*{fsge} approach allows directional stimuli for $\stimw$ by replacing the L2-norm of the \gls*{wss} vector $\vc{\tau}_w$ in \eqref{eq_stimuli} by a directionality measure like transWSS \eqref{eq_transwss}}. In the future, incorporating local and directional \gls*{wss} will enable the testing of hypotheses in \gls*{fsge}-mediated aneurysm growth and atherosclerotic lesion formation using \gls*{fsge}. Since current \gls*{gr} models were designed with the static and scalar nature of reduced fluid dynamics in mind, we expect future biological insights from \gls*{fsge} to arise from the advancements outlined in this work.

\section{Acknowledgements}
This work was supported by NIH Grants K99HL161313, R01HL139796, R01HL159954, the Stanford Maternal and Child Health Research Institute, and the Additional Ventures Foundation Cures Collaborative. We further thank Dr. Richard Schussnig for his help with IQN-ILS and Drs. Alice Cortinovis and Luca Pegolotti for their help with linear solvers. In addition, we thank Drs. Yuri Bazilevs, Fabian Br\"au, David Li, Matteo Salvador, and Vijay Vedula for stimulating discussions on the stability of the \gls*{cmme} model.

\newpage

\bibliographystyle{elsarticle-num} 
\bibliography{references,software}

\end{document}